\documentclass[letterpaper]{article}
\usepackage{aaai2026} 
\usepackage{times} 
\usepackage{helvet} 
\usepackage{courier} 
\usepackage[hyphens]{url} 
\usepackage{graphicx} 
\usepackage{graphicx}  
\usepackage{natbib}  
\usepackage{caption}  
\usepackage{algorithm}
\usepackage{algorithmic}

\usepackage{newfloat}
\usepackage{listings}
\DeclareCaptionStyle{ruled}{labelfont=normalfont,labelsep=colon,strut=off} 
\floatstyle{ruled}
\newfloat{listing}{tb}{lst}{}
\floatname{listing}{Listing}

\usepackage{amsmath}
\usepackage{amsthm}
\usepackage{tabularx}
\usepackage[titletoc,title]{appendix}

\theoremstyle{plain}
\newtheorem{theorem}{Theorem}
\newtheorem*{theorem*}{Theorem}

\newtheorem*{claim*}{Claim}

\newtheorem*{lemma*}{Lemma}

\newtheorem*{proposition*}{Proposition}

\theoremstyle{remark}

\usepackage{mathtools}
\usepackage{bm}
\usepackage{amssymb}
\usepackage{amsthm}
\usepackage{xcolor, soul}

\newcommand{\cov}{\textnormal{cov}}
\newcommand{\lbl}{\textnormal{lbl}}
\newcommand{\ppi}{\textnormal{PPI}}
\newcommand{\ppipp}{\textnormal{PPI++}}
\newcommand{\reppi}{\textnormal{RePPI}}
\newcommand{\aipw}{\textnormal{AIPW}}

\newcommand{\indic}{\bm{1}}

\newcommand{\cf}{\mathcal{F}}

\newcommand{\ci}{\mathcal{I}}

\newcommand{\cs}{\mathcal{S}}

\newcommand{\nophoto}{\textnormal{nophoto}}

\renewcommand{\leq}{\leqslant}

\newcommand{\e}{\mathbb{E}}
\newcommand{\real}{\mathbb{R}}

\newcommand{\var}{\mathrm{Var}}

\DeclareMathOperator*{\argmin}{arg\,min}

\title{Scalable Vision-Guided Crop Yield Estimation}
\author{
    Harrison H. Li\textsuperscript{\rm 1}\equalcontrib,
    Medhanie Irgau\textsuperscript{\rm 2}\equalcontrib,
    Nabil Janmohamed\textsuperscript{\rm 3},
    Karen Solveig Rieckmann\textsuperscript{\rm 3},
    David B. Lobell\textsuperscript{\rm 4}
}
\affiliations{
    \textsuperscript{\rm 1}Department of Mathematics, Harvey Mudd College, Claremont, CA USA \\

    \textsuperscript{\rm 2}Department of Computer Science, Stanford University, Stanford, CA USA \\
    \textsuperscript{\rm 3}Pula Advisors AG, Mollis, Switzerland\\
    \textsuperscript{\rm 4}Department of Earth System Science and Center on Food Security and the Environment \\
    Stanford University, Stanford, CA USA \\
    harrli@hmc.edu,
    mirgau@stanford.edu,
    nabil@pula.io,
    krieckmann@pula.io,
    dlobell@stanford.edu
}

\nocopyright

\begin{document}
\maketitle
\begin{abstract}
Precise estimation and uncertainty quantification for average crop yields are 
critical for agricultural monitoring and decision making. Existing data collection methods, such as crop cuts in randomly sampled fields at harvest time, are relatively time-consuming. Thus, we propose an approach based on prediction-powered inference (PPI) to supplement these crop cuts with less time-consuming field photos.
After training a computer vision model to predict the ground truth crop cut yields from the photos,
we learn a ``control function" that recalibrates these predictions with the spatial coordinates of each field. This enables fields with photos but not crop cuts to be leveraged to improve the precision of zone-wide average yield estimates.
Our control function is learned by training on a dataset of nearly 20,000 real crop cuts and photos of rice and maize fields in sub-Saharan Africa.
To improve precision, we pool training observations across different zones within the same first-level subdivision of each country.
Our final PPI-based point estimates of the average yield are provably asymptotically unbiased and cannot increase the asymptotic variance beyond that of the natural baseline estimator --- the sample average of the crop cuts ----- as the number of fields grows.
We also propose a novel bias-corrected and accelerated (BCa) bootstrap to construct accompanying confidence intervals.
Even in zones with as few as 20 fields,
the point estimates show significant empirical improvement over the baseline,
increasing the effective sample size by as much as 73\% for rice and by 12-23\% for maize.
The confidence intervals are accordingly shorter at minimal cost to empirical finite-sample coverage.
This demonstrates the potential for relatively low-cost images to make area-based crop insurance more affordable and thus spur investment into sustainable agricultural practices.
\end{abstract}

\begin{links}
    \link{Code}{https://github.com/medhanieirgau/scalable-vision-guided-crop-yield-estimation}
    \link{Datasets}{https://doi.org/10.5281/zenodo.17626117}
\end{links}

\section{Introduction}

Improved agricultural productivity is critical to achieve greater food security and economic growth worldwide. A significant impediment to progress in many countries has been the lack of reliable data on cropping outcomes such as yield, which can impede many forms of investments. A prime example is that farmers often lack access to affordable and effective crop insurance, which would enable higher levels of investment and risk-taking needed to raise productivity. Effective insurance, in turn, relies on the ability to accurately assess cropping conditions and yields across large and heterogeneous cropping regions. Historically, this information has proven to be difficult and costly to obtain.

Recent progress has been made by conducting systematic large-scale crop cutting measurements during the harvest season. These crop cuts are then averaged over pre-defined regions called zones to provide an objective measure of average yields that forms the basis for area-based insurance products.

The motivation for the present work comes from the promise of photographs or aerial imagery as more affordable and scalable approaches to yield estimation~\citep{khaki2021deepcorn,li2022maize}. While they have commonly been promoted as potential alternatives to ground-based measures like crop cuts, they often explain only half or less of true yield variability in complex smallholder farming settings \cite{darra2023can} and can be biased~\citep{lobell2020eyes}.
This makes using such yield proxies in the place of ground measurements inadequate for insurance and re-insurance providers. 
We thus seek to instead supplement ground measurements with images --- specifically, photos of fields taken at harvest time --- to increase the effective sample size without introducing bias.

To do so,
we propose leveraging recent statistical methods to efficiently combine small amounts of expensive, error-free data with larger amounts of cheaper but less accurate measures.
After training a computer vision model to predict yields from photos,
we post-process these predictions by adopting ideas from the framework of prediction-powered inference (PPI).
Importantly, we are able to guarantee that in large samples, our zone-level average yield estimates remain unbiased and cannot have larger variance than using the crop cuts alone,
regardless of the performance of the computer vision model.
However, a better computer vision model enables a lower asymptotic variance for the final estimator,
to a degree that we can quantify.
We empirically validate these theoretical guarantees under relatively small sample sizes using an extensive dataset from Pula, a provider of area-based crop insurance products in sub-Saharan Africa.



The remainder of the paper is organized as follows. Section~\ref{sec:data} describes the dataset and preprocessing steps. Section~\ref{sec:ppi} reviews the prediction-powered inference framework and introduces our estimation approach. Section~\ref{sec:photo_prediction_model} details the photo-based yield prediction models underlying our methodology, and Section~\ref{sec:cross_zones} presents some innovations that leverage the nature of area-based insurance data
to improve learning of the control function that underpins PPI.
Section~\ref{sec:results} reports empirical findings and Section~\ref{sec:discussion} concludes.

\section{Data} \label{sec:data}
\begin{figure}[!t]
\centering
\includegraphics[width=0.45\linewidth]{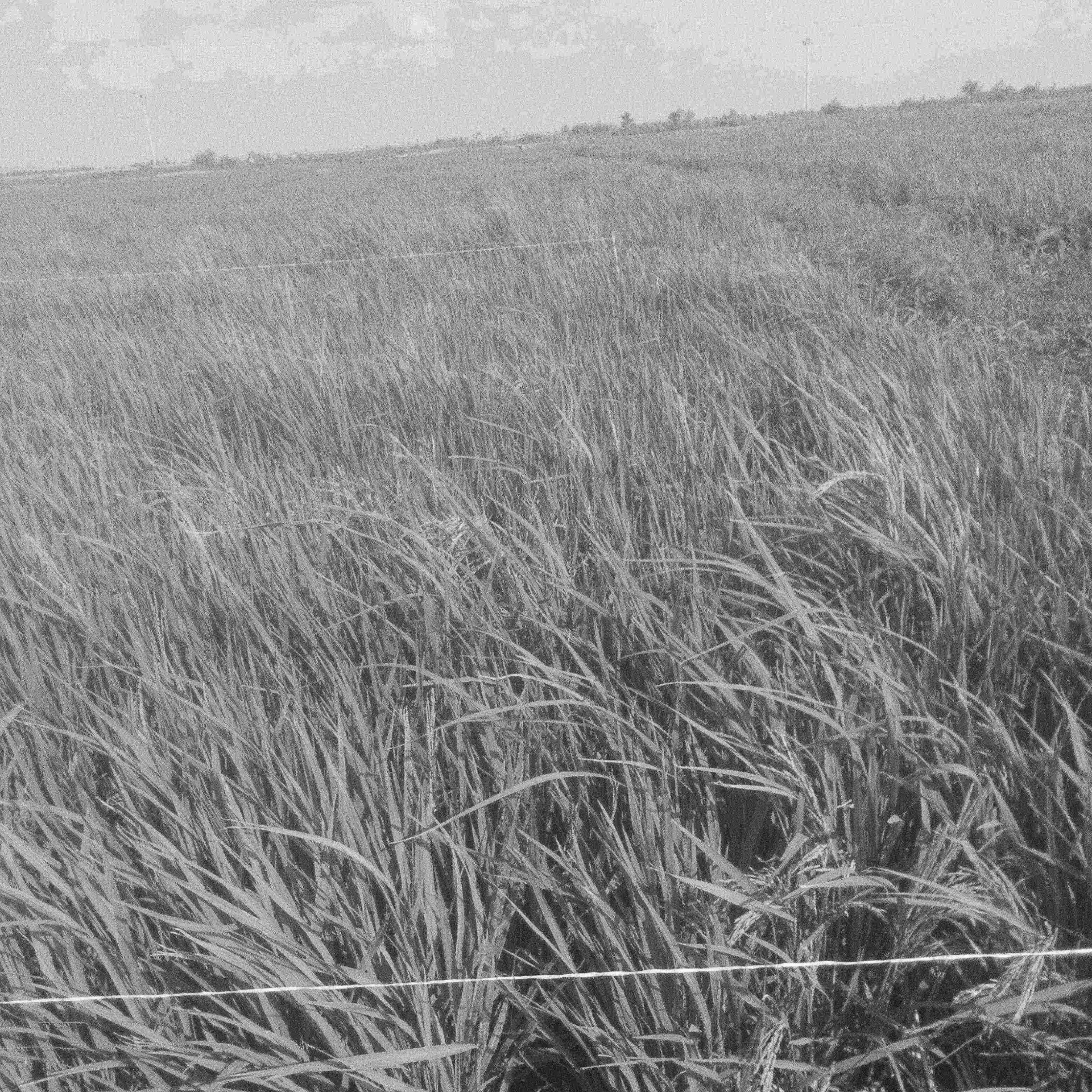}
\includegraphics[width=0.45\linewidth]{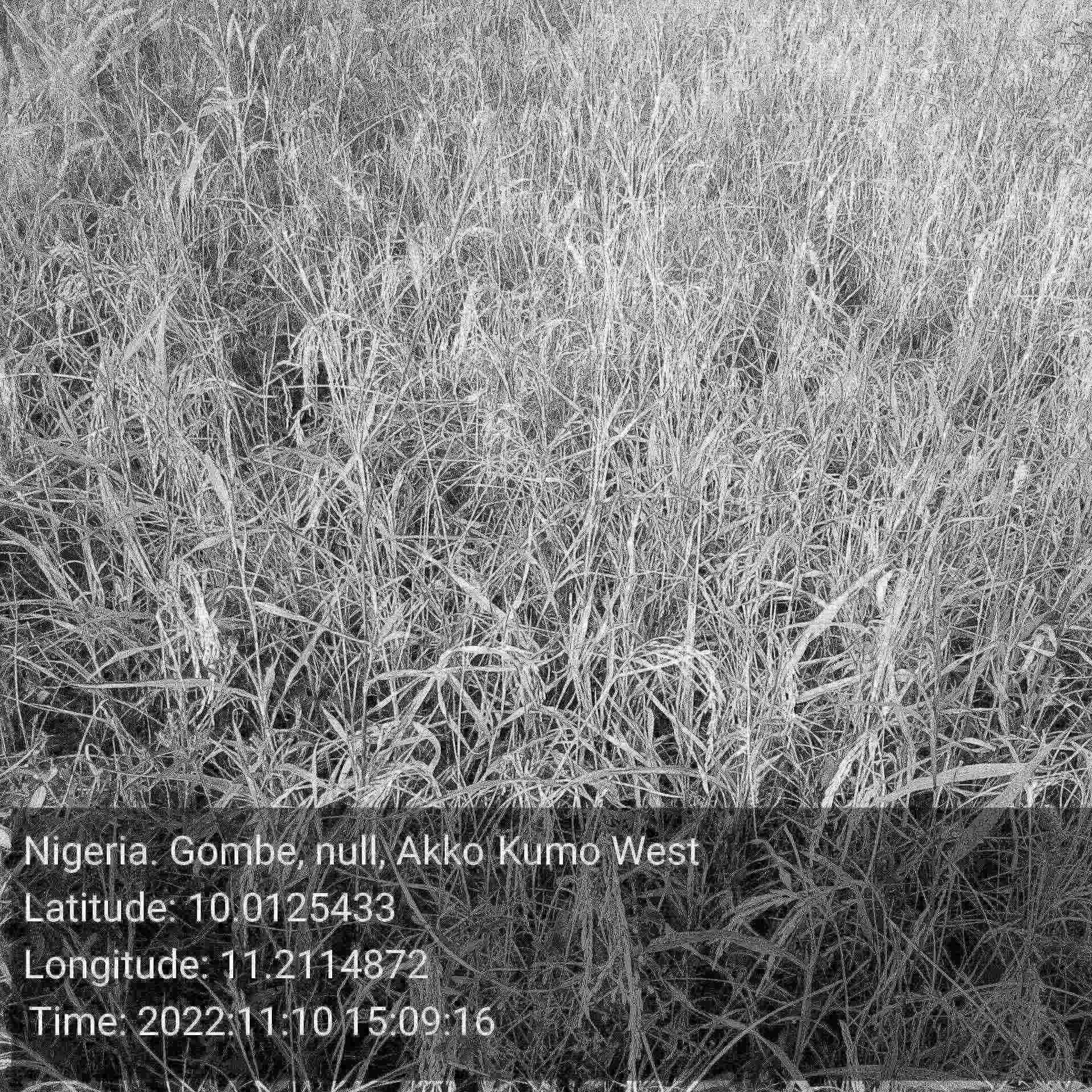}
\includegraphics[width=0.45\linewidth]{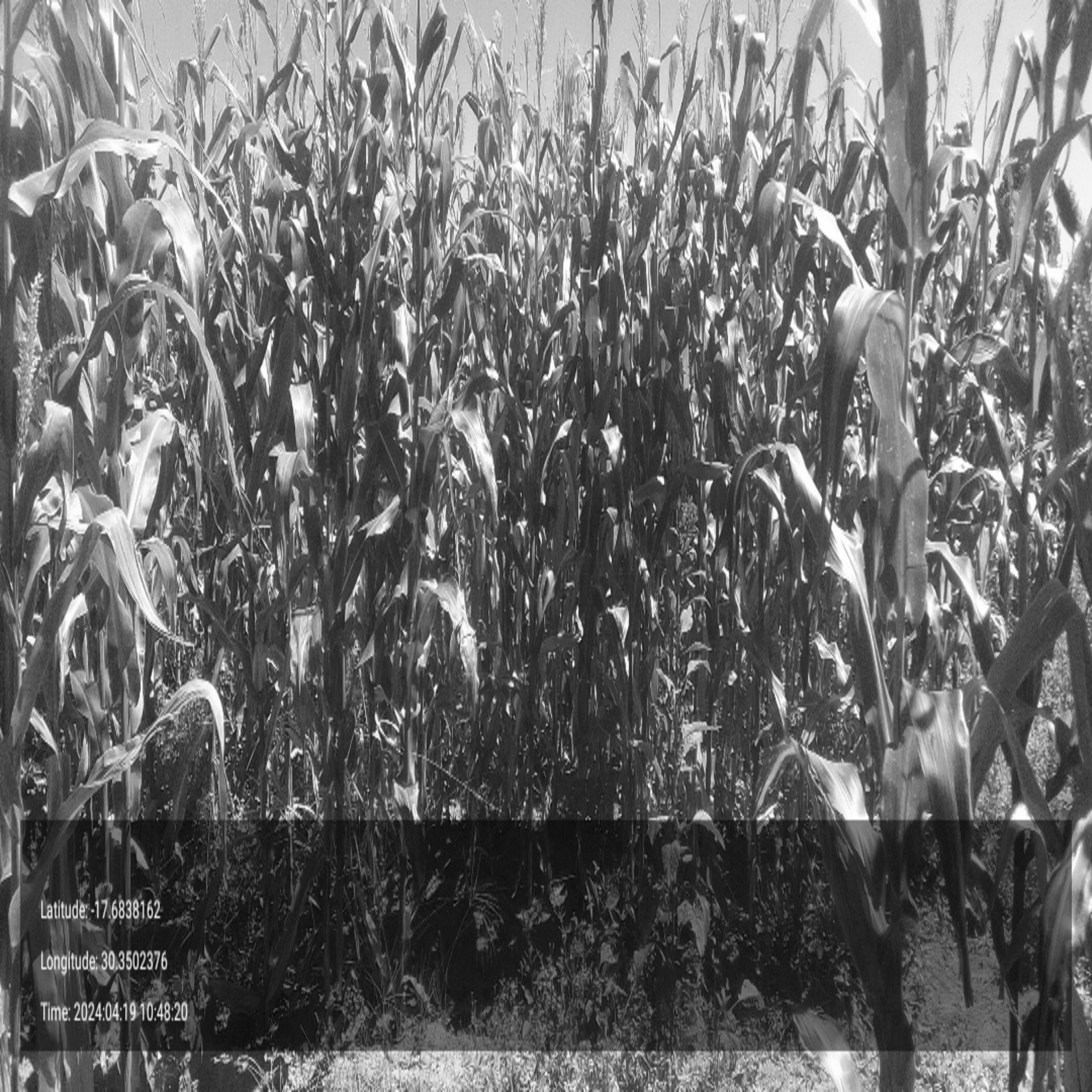}
\includegraphics[width=0.45\linewidth]{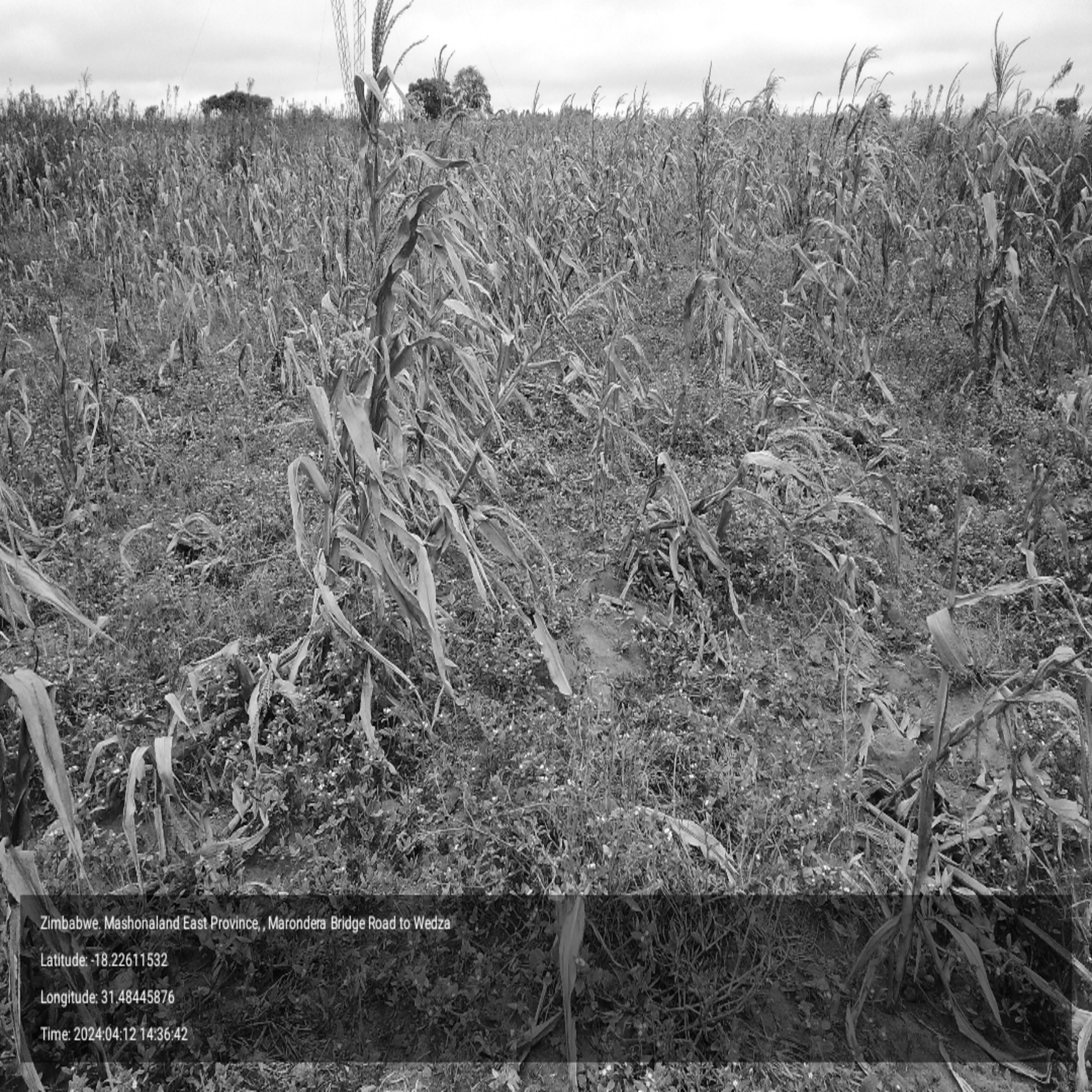} \\
\caption{(Top row) Sample rice field photos from Nigeria in harvest year 2022.
The fields have ground truth (crop cut) yields of $4.7$ $\texttt{mt ha}^{-1}$ (left) and $0.0088$ $\texttt{mt ha}^{-1}$ (right).
(Bottom row) Sample maize field photos from Zimbabwe in harvest year 2024.
The fields have ground truth (crop cut) yields of $9.1$ $\texttt{mt ha}^{-1}$ (left) and $0.0061$ $\texttt{mt ha}^{-1}$ (right).}
\label{fig:sample_photos}
\end{figure}

Our dataset contains field-level observations from four (country, harvest year) pairs: rice fields from (Nigeria, 2022) and maize (corn) fields from (Zambia, 2023), (Zambia, 2024), and (Zimbabwe, 2024).
For each field, 
two random crop cuts were taken at harvest time.
Their average is treated as the ground truth yield,
in units of $\texttt{mt ha}^{-1}$.
Photographs were also manually captured for each field.
Some examples
are provided in Fig.~\ref{fig:sample_photos}.
The data from Zambia are commercially sensitive;
however, we provide the complete data from Nigeria and Zimbabwe in the link following the abstract of this paper.
Currently, we do not have access to photos from additional fields;
in our evaluations we use bootstrapping to simulate the desired future setting where there are additional ``unlabeled" fields with photos but not crop cuts (Section~\ref{sec:results}).

Each field is assigned to a unique zone for area-based insurance purposes.
Since we are interested in improving estimates of zone-level mean yields,
to reduce noise in our performance metrics we filter our dataset to only zones with at least 20 observations.
Table~\ref{table:zone_summaries} provides summary statistics for the final filtered dataset.

Prior to public release, we manually reviewed all $\sim$7,000 field photos in Nigeria and Zimbabwe to ensure that no identifiable individuals or other sensitive content were present. Approximately 80 images containing identifiable people were removed. The released dataset therefore excludes all such content.

\begin{figure}[!t]
\centering
\includegraphics[width=\linewidth]{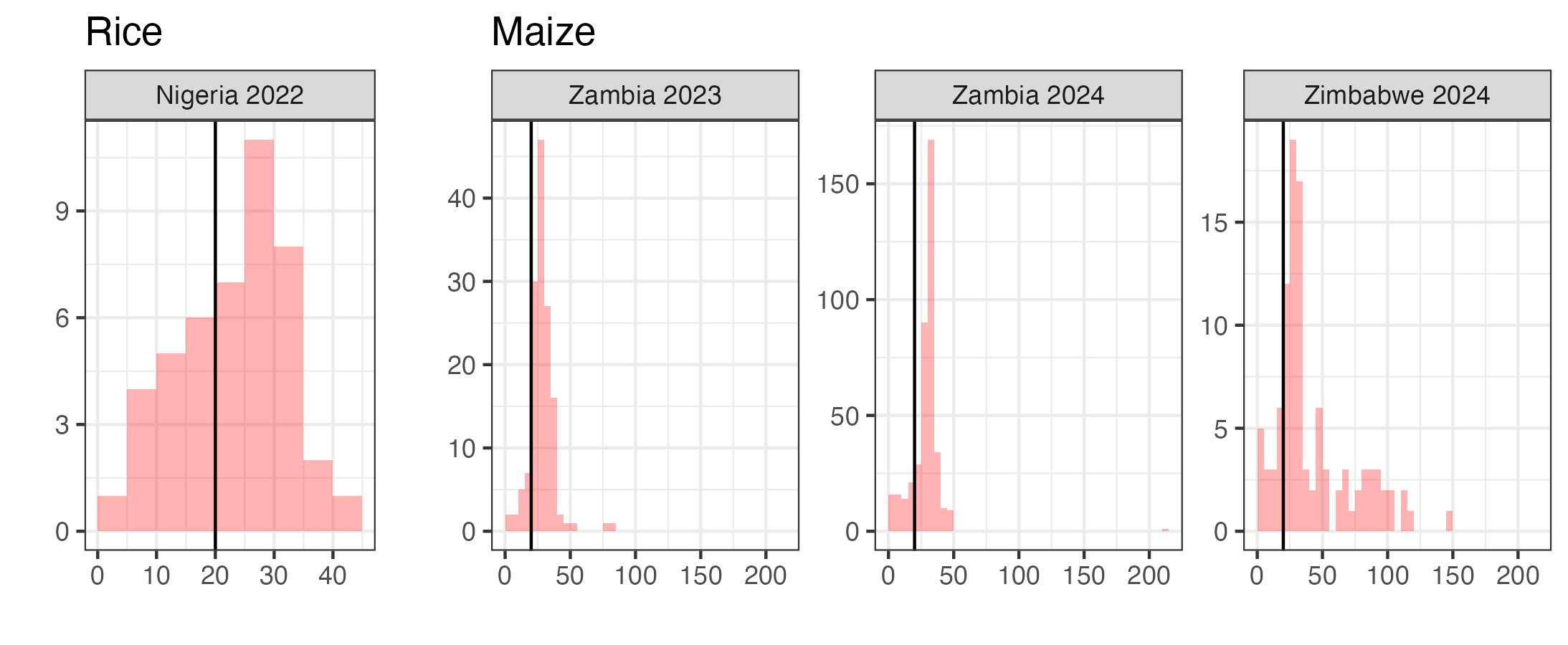}
\caption{Histograms of the number of fields within each zone in our dataset, separated by country and harvest year}
\label{fig:zone_summaries}
\end{figure}

\begin{table}[!t]
\centering
\caption{The number of eligible zones (i.e., zones with at least 20 fields) and the total number of fields across the eligible zones in the dataset by country and harvest year
}
\label{table:zone_summaries}
\begin{tabular}{|c|c|c|c|c|}
\hline
Crop & Country & Harvest year & Zones & Fields \\
\hline
Rice & Nigeria & 2022 & 29 & 826 \\
Maize & Zambia & 2023 & 126 & 3759 \\
Maize & Zambia & 2024 & 342 & 10727 \\
Maize & Zimbabwe & 2024 & 87 & 4173 \\
\hline
\end{tabular}
\end{table}

\section{Prediction-powered inference}\label{sec:ppi}
We now provide some background on prediction-powered inference (PPI)
and show the ideas can be leveraged in our context to effectively combine field photos with crop cuts for mean yield estimation.

For clarity,
we begin by considering the data in a single zone,
although we will end up pooling data from multiple zones to improve the finite sample performance of our final zone-level average yield estimates
(Section~\ref{sec:cross_zones}).
The data in a zone consists of ``labeled" fields $i=1,\ldots,n$ where we observe the ground truth yields (``labels") $Y_i$, 
photos $V_i$,
and covariates $X_i$,
along with ``unlabeled" fields $i=n+1,\ldots,n+N$ for which only the photos and covariates are available.
As noted in Section~\ref{sec:data},
there are no unlabeled samples in our dataset,
but we present our methodology in anticipation of a future application with additional unlabeled samples that can be cheaply collected.
For our evaluations,
we only use latitude and longitude --- readily available in our dataset --- as our two covariates $X_i$,
though future implementations could incorporate more covariates.
The estimand of interest is $\theta = \mathbb{E}[Y_i]$.
We access the information in the photos $V_i$ through predictions $\hat{Y}_i=g(V_i)$ from a computer vision model $g$.
We defer the details of how to train $g(\cdot)$ to Section~\ref{sec:photo_prediction_model};
for now, we treat it as a black box.

\begin{algorithm}[!t]
\caption{\texttt{PPI++}}
\label{alg:estimator}
\begin{algorithmic}[1]
\REQUIRE Labeled data $\{(Y_i,W_i)\}_{i=1}^n$,
unlabeled data $\{W_i\}_{i=n+1}^{n+N}$,
control function $f(\cdot)$
\STATE $\hat{\theta}_{\lbl} \leftarrow \frac{1}{n} \sum_{i=1}^{n} Y_i$
\STATE $\bar{f}_n \leftarrow \frac{1}{n} \sum_{i=1}^n f(W_i)$
\STATE $\bar{f}_N \leftarrow \frac{1}{N} \sum_{i=n+1}^{n+N} f(W_i)$
\STATE $\bar{f} \leftarrow \frac{1}{n+N} \sum_{i=1}^{n+N} f(W_{i})$
\STATE $\widehat{\cov}(Y, f(W)) \leftarrow \frac{1}{n-1} \sum_{i=1}^{n} (Y_{i}-\hat{\theta}_{\lbl})(f(W_{i})-\bar{f}_{n})$, $\quad \widehat{\var}(f(W)) \leftarrow \frac{1}{n+N-1} \sum_{i=1}^{n+N} (f(W_{i})-\bar{f})^2$
\STATE $\hat{\lambda} \leftarrow \frac{N}{n+N} \frac{\widehat{\cov}(Y,f(W))}{\widehat{\var}(f(W))}$
\RETURN $\hat{\theta}_{\lbl} - \hat{\lambda} \left(\bar{f}_{n} - \bar{f}_{N} \right)$
\end{algorithmic}
\end{algorithm}

Algorithm~\ref{alg:estimator} details how to compute the proposed PPI++ estimator $\hat{\theta}_{\ppipp}$ for $\theta$ given these data.
Following~\citet{angelopoulos2023ppi++},
the PPI++ estimator takes the form
\begin{equation}
\label{eq:ppipp}
\hat{\theta}_{\ppipp} = \hat{\theta}_{\lbl} - \hat{\lambda} \left(\frac{1}{n} \sum_{i=1}^n f(W_i) - \frac{1}{N} \sum_{i=n+1}^{n+N} f(W_i) \right)
\end{equation}
where $\hat{\theta}_{\lbl} = n^{-1} \sum_{i=1}^n Y_i$
is simply the sample average of the yields of the labeled fields.
The additional components of the PPI++ estimator~\eqref{eq:ppipp}
are the control function $f$ ---
designed to predict the yields $Y$ from the combination $W=(\hat{Y},X)$ of photo model predictions and covariates ---
and the coefficient $\hat{\lambda}$,
which can depend on $f$.
If the observations in the labeled and unlabeled fields are independent and drawn from the same distribution,
then provided $\hat{\lambda}$ converges to a non-random limit,
the PPI++ estimator $\hat{\theta}_{\ppipp}$ is asymptotically unbiased for $\theta$ for any control function $f$.
While this assumption can be restrictive in some applications,
an area-based crop insurer can ensure it holds by design by choosing which fields receive crop cuts uniformly at random.
We thus assume it holds throughout this paper.

~\citet{angelopoulos2023ppi++} showed that given any square integrable control function $f$,
the coefficient
\begin{equation}
\label{eq:lambda_hat}
\hat{\lambda} = \frac{N}{n+N} \frac{\widehat{\cov}(Y_i, f(W_i))}{\widehat{\var}(f(W_i))}
\end{equation}
where $\widehat{\cov}$ and $\widehat{\var}$ denote the usual sample-based estimates of covariance and variance,
respectively,
minimizes the asymptotic variance of $\hat{\theta}_{\ppipp}$ as the sample sizes $n$ and $N$ grow.
The choice of $\hat{\lambda}$ in~\eqref{eq:lambda_hat} ensures
the estimator $\hat{\theta}_{\ppipp}$ asymptotically dominates (i.e., cannot have higher asymptotic variance than) $\hat{\theta}_{\lbl}$ which ignores the photos and corresponds to setting $\hat{\lambda}=0$ in~\eqref{eq:ppipp},
along with the original PPI estimator of~\citet{angelopoulos2023prediction} which corresponds to setting $\hat{\lambda}=1$ in~\eqref{eq:ppipp}.
While recent work by~\citet{ji2025predictions} has found that a ``recalibrated PPI" (RePPI) estimator can further asymptotically dominate PPI++,
we show in Appendix~\ref{app:reppi} that RePPI is in fact equivalent to our PPI++ estimator in our setting.

It remains to make a prudent choice of the control function $f$.
Let $\mu(w) = \e[Y \mid W=w]$
be the conditional mean outcome given the photo predictions and covariates (i.e., $W$).
When $f=\mu$ we have $\hat{\lambda}=N/(n+N)$ from~\eqref{eq:lambda_hat}.
Plugging these choices of $(f,\hat{\lambda})$
into~\eqref{eq:ppipp} yields the augmented inverse propensity weighted (AIPW) estimator $\hat{\theta}_{\aipw}$,
which is known to be semiparametrically efficient for this problem~\citep{robins1994estimation, angelopoulos2023ppi++}.
This means that picking $(f,\hat{\lambda})=(\mu,N/(n+N))$ minimizes the asymptotic variance not only among all estimators of the form~\eqref{eq:ppipp},
but also among all regular, asymptotically linear estimators for $\theta$.

However, in reality $\mu$ is not known
so we cannot simply ``pick" $f=\mu$.
Naturally, we can take $f$ to be a data-driven estimate $\hat{\mu}$ of $\mu$,
learned via some supervised learning algorithm to predict $Y$ from $W$.
Semiparametric efficiency, however, would require
this estimated $\hat{\mu}$ to be asymptotically equivalent to (i.e., consistent in mean square for) $\mu$.
With $n$ generally on the order of dozens (Fig.~\ref{fig:zone_summaries}),
it is not prudent to assume that $\mu$ can be well approximated nonparametrically.
This is an important argument in favor of using the PPI++ estimator in lieu of the AIPW estimator.
The PPI++ and AIPW estimators differ in their choice of $\hat{\lambda}$.
Whereas AIPW sets $\hat{\lambda}=N/(n+N)$ based on the assumption that $f \approx \mu$,
the PPI++ estimator adaptively chooses $\hat{\lambda}$ via~\eqref{eq:lambda_hat} to approximate the optimal choice for the $f$ actually learned.
The efficiency of AIPW,
however,
still incentivizes learning a control function $f$ that is as close to $\mu$ as possible.

\begin{figure}[!t]
\centering
\includegraphics[width=\linewidth]{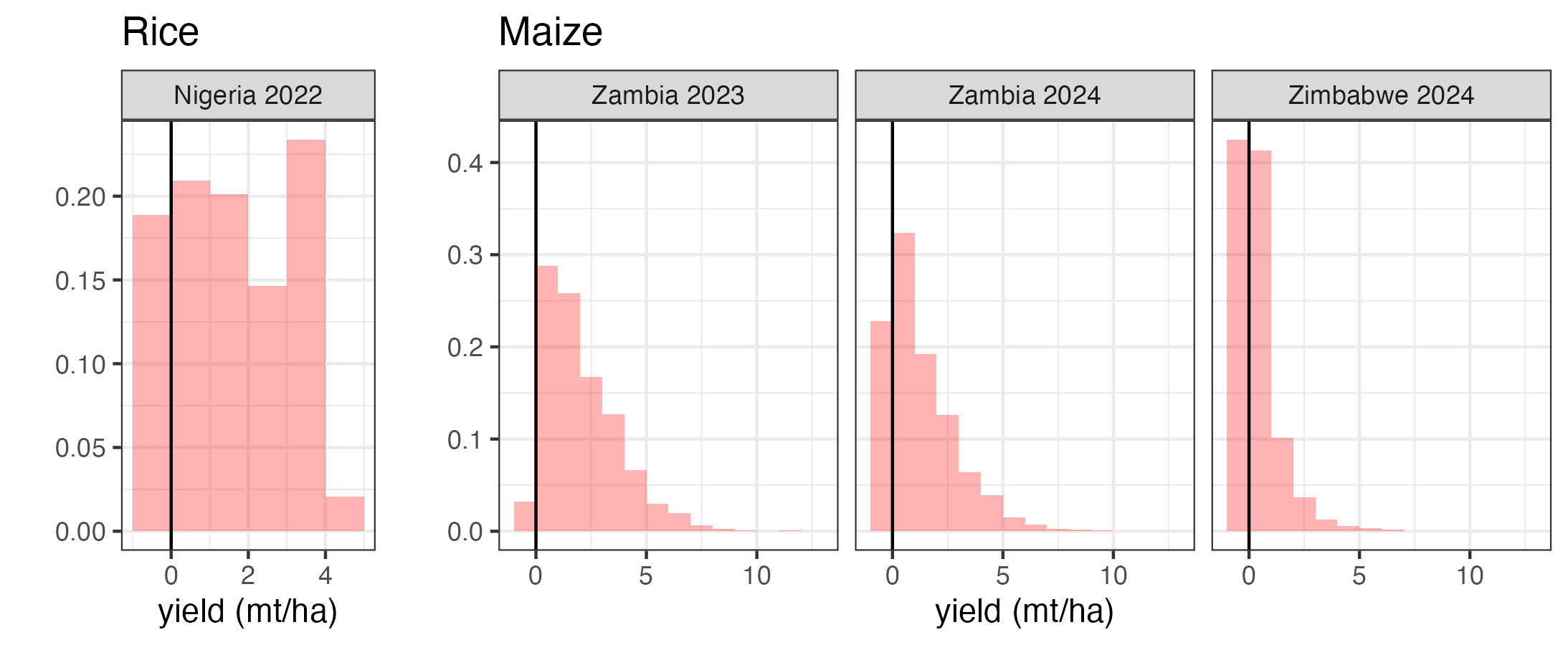} \\
\caption{Histograms of field-level yields, separated by country and harvest year.
All observations to the left of the solid vertical lines at 0 are exact zeros.}
\label{fig:yields}
\end{figure}

\begin{algorithm}[!t]
\caption{\texttt{PPBootBCa}}
\label{alg:bootstrap}
\begin{algorithmic}[1]
\REQUIRE Labeled data $\{(Y_i,W_i)\}_{i=1}^n$,
unlabeled data $\{W_i\}_{i=n+1}^{n+N}$,
control function $f(\cdot)$,
significance level $\alpha$,
number of bootstrap samples $B$
\STATE Compute point estimate \\
$\hat{\theta}_{\ppipp} \leftarrow \texttt{PPI++}\left(\{(Y_i,W_i)\}_{i=1}^n, \{W_i\}_{i=n+1}^{n+N}, f(\cdot)\right)$
\FOR{$b=1,\ldots,B$}
\STATE Get bootstrap samples $\left\{(Y_i^{(b)},W_i^{(b)})\right\}_{i=1}^n$, $\left\{W_i^{(b)}\right\}_{i=n+1}^{n+N}$ by sampling with replacement
\STATE Compute bootstrap point estimate $\hat{\theta}_{\ppipp}^{(b)} \leftarrow \texttt{PPI++}\left(\left\{(Y_i^{(b)},W_i^{(b)})\right\}_{i=1}^n, \left\{W_i^{(b)}\right\}_{i=n+1}^{n+N}, f(\cdot)\right)$ 
\ENDFOR
\STATE $z_0 \leftarrow \Phi^{-1}\left(B^{-1}\sum_{b=1}^B \indic\left[\hat{\theta}_{\ppipp}^{(b)} \leq \hat{\theta}_{\ppipp}\right] \right)$ \hfill \textit{// Bias-correction parameter; $\Phi$ is $\mathcal{N}(0,1)$ CDF}
\STATE $z_L \leftarrow z_0 + \Phi^{-1}(\alpha/2), \quad z_U \leftarrow z_0 + \Phi^{-1}(1-\alpha/2)$
\FOR{$m=1,\ldots,n+N$}
\STATE $\hat{\theta}^{(-m)} \leftarrow \texttt{PPI++}\bigg(\{(Y_i,W_i)\}_{i\in \{1,\ldots,n\} \setminus \{m\}}$, $\{W_i\}_{i \in \{n+1,\ldots,n+N\} \setminus \{m\}}, f(\cdot)\bigg)$
\ENDFOR
\STATE $\bar{\theta} \leftarrow (n+N)^{-1} \sum_{m=1}^{n+N} \hat{\theta}^{(-m)}$
\STATE $u \leftarrow [\bar{\theta}-\hat{\theta}^{(-1)},\ldots,\bar{\theta}-\hat{\theta}^{(-[n+N])}]$
\STATE $\gamma \leftarrow  \frac{\sum_{m=1}^{n+N} u_m^3}{6(\sum_{m=1}^{n+N} u_m^2)^{3/2}}$ \hfill \textit{// Acceleration parameter}
\RETURN $\bigg[\texttt{quantile}\left(\{\hat{\theta}_{\ppipp}^{(b)}\}_{b=1}^B, \Phi\left(z_0+\frac{z_L}{1-\gamma z_L}\right)\right),$ $\texttt{quantile}\left(\{\hat{\theta}_{\ppipp}^{(b)}\}_{b=1}^B, \Phi\left(z_0+\frac{z_U}{1-\gamma z_U}\right)\right)\bigg]$
\end{algorithmic}
\end{algorithm}

We also seek the quantify the uncertainty in our point estimate $\hat{\theta}_{\ppipp}$.
While $\hat{\theta}_{\ppipp}$ satisfies a central limit theorem
(e.g., Theorem~\ref{thm:main} below),
enabling asymptotically valid confidence intervals based on the normal distribution as $n,N \rightarrow \infty$,
ground truth yields tend to be substantially skewed and zero-inflated,
particularly for maize (Fig.~\ref{fig:yields}),
so such intervals do not attain the desired coverage levels for small $n$.
To ameliorate these issues,
in Algorithm~\ref{alg:bootstrap}
we propose a nonparametric bias-corrected and accelerated (BCa) bootstrap~\citep{efron1987better} to generate confidence intervals for $\theta$.
Confidence intervals from the BCa bootstrap satisfy desirable second-order asymptotic coverage properties and are designed to correct for issues like skewness in the data distribution~\citep{efron2021computer}.
Algorithm~\ref{alg:bootstrap} can be seen as a variant of the prediction-powered bootstrap \texttt{PPBoot}~\citep{zrnic2024note}.
We find empirically that our BCa bootstrap exhibits superior performance over an alternative based on the bootstrap-$t$ method
as well as over both standard asymptotic confidence intervals based on the central limit theorem (CLT) and the percentile bootstrap intervals suggested by \texttt{PPBoot} (Appendix~\ref{app:ci_type_compare}).

\begin{algorithm}[!t]
\caption{Full procedure}
\label{alg:full_pipeline}
\begin{algorithmic}[1]
\REQUIRE Ground truth yields $\{Y_{ji}\}_{i=1}^{n_j}$,
photos $\{V_{ji}\}_{i=1}^{n_j+N_j}$,
and covariates $\{X_{ji}\}_{i=1}^{n_j+N_j}$ for zones $j=1,\ldots,J$;
number of cross-fitting folds $K$
\FOR{zone $j=1,\ldots,J$}
\STATE Partition indices $(j,1),\ldots,(j,n_j)$ randomly and as evenly as possible into folds $I_{j1},\ldots,I_{jK}$
\STATE $\ci_k \leftarrow \ci_k \cup I_{jk}, \quad k=1,\ldots,K$
\ENDFOR
\STATE $\ci \leftarrow \cup_{k=1}^K \ci_k$
\FOR{fold $k=1,\ldots,K$} 
\STATE Train photo model $\hat{g}^{(-k)}(\cdot)$ to predict $Y_{ji}$ from $V_{ji}$ using all observations with $(j,i) \in \ci \setminus \ci_k$
\STATE Compute photo predictions on labeled observations: $\hat{Y}_{ji} \leftarrow \hat{g}^{(-k)}(V_{ji}), \quad (j,i) \in \ci_k$ 
\ENDFOR
\STATE Compute photo predictions on unlabeled observations by averaging all $K$ models: \\
$\hat{Y}_{ji} \leftarrow \frac{1}{K} \sum_{k=1}^K \hat{g}^{(-k)}(V_{ji}), \quad j=1,\ldots,J, \quad i=n_j+1,\ldots,n_j+N_j$
\STATE $W_{ji} \leftarrow (\hat{Y}_{ji},X_{ji}), \quad \psi(W_{ji}) \leftarrow (1,\hat{Y}_{ji},X_{ji})', \quad j=1,\ldots,J, \quad i=1,\ldots,n_j+N_j$
\STATE Partition zones $j=1,\ldots,J$ into $R$ sets $\cs_1,\ldots,\cs_R$ by study region
\FOR{study region $r=1,\ldots,R$}
\STATE Learn coefficients $\hat{\beta}_r$ via 5-fold cross-validated LASSO to predict $Y$ from $\psi(W)$ using all labeled observations in $\cs_r$
\STATE $f_r(\cdot) \leftarrow \hat{\beta}_r^{\top}\psi(\cdot)$
\ENDFOR
\STATE \texttt{estimates} $\leftarrow []$, \texttt{CIs} $\leftarrow []$
\FOR{$r=1,\ldots,R$}
\FOR{$j \in \cs_r$}
\STATE $\hat{\theta}_{\ppipp, j} \leftarrow \texttt{PPI++}\bigg(\{(Y_{ji},W_{ji})\}_{i=1}^{n_j},$ \\
\quad $\{W_{ji}\}_{i=n_j+1}^{n_j+N_j}, f_r(\cdot)\bigg)$
\STATE \texttt{estimates.append} $\big(\hat{\theta}_{\ppipp,j}\big)$
\STATE \texttt{CIs.append(} $\texttt{PPBootBCa}\big(\{(Y_{ji},W_{ji})\}_{i=1}^{n_j},$ \\ 
$\qquad \{W_{ji}\}_{i=n_j+1}^{n_j+N_j},$ $f_r(\cdot)\big)$
\ENDFOR
\ENDFOR
\RETURN \texttt{estimates}, \texttt{CIs}
\end{algorithmic}
\end{algorithm}

\section{Computer vision models}\label{sec:photo_prediction_model}

Our vision models are convolutional neural networks based on the ResNet-50 architecture~\citep{he2015deepresiduallearningimage}, pretrained on ImageNet~\citep{russakovsky2015imagenetlargescalevisual} and fine-tuned to predict the ground truth maize or rice yield from crop cuts. 
To adapt each model for regression, we replace the final classification layer with a fully connected layer that outputs a single scalar representing the predicted yield. 
Training minimizes mean squared error (MSE) using the Adam optimizer.

We aim to learn a separate model for each (country, harvest year) pair.
To ensure robust evaluation and avoid overfitting, we adopt a parallelized 5-fold cross-fitting approach. 
That is, for each (country, harvest year) pair,
labeled data within each zone is split randomly into five folds and five models are trained in parallel.
Each of these models is trained on the observations across all zones from four of the folds and then used to predict the crop cut yields from the photos in the held-out fold.
Then each field has a predicted yield from exactly one of the five models,
which was trained on the folds not containing that field.

At the end of an epoch, we aggregate such predictions to form a full set of validation predictions and compute an evaluation metric: the average within-zone \(R^2\) score, computed from held-out predictions. 
Specifically, we compute the \(R^2\) between predicted and true yields separately within each zone, 
and then take a simple average across all zones in the (country, harvest year) pair. 
This metric is sensitive to how well the model captures yield variation within individual zones, and directly governs the amount variance reduction we can expect from our methods (Section~\ref{sec:cross_zones}).
Through training, we track the five model checkpoints with the highest validation average within-zone \(R^2\) (one for each fold) and save these to make our final predictions.
The within-zone $R^2$ estimates for these models are given for each (country, year) pair in Table~\ref{table:model_r2s}.
For reference,
we also provide the substantially higher ``cross-zone" $R^2$ estimates based on predictions across all zones.

We trained all models using the Adam optimizer for 10 epochs with an initial learning rate selected from a sweep over {1e-4, 2e-4, 3e-4, 5e-4, 1e-3}, choosing the best based on validation average within-zone $R^2$ for each (country, harvest year) pair. A learning rate of 2e-4 was selected for (ZM, 2023) and (ZW, 2024), while a learning rate of 3e-4 was selected for (ZM, 2024) and (NG, 2022). All experiments were run on a cluster with 20 CPU cores, 20 GB RAM, and 5 NVIDIA A4000 GPUs, running Ubuntu 20.04.6 LTS. We used a batch size of 128 during training.
\begin{table}[!t]
    \centering
    \caption{$R^2$ scores of our photo model predictions per (country, harvest year) pair. The within-zone numbers are simple averages across all zones.}
    \label{table:model_r2s}
    \begin{tabularx}{\linewidth}{c|X|X|X|X}
        \hline
         $R^2$ & (NG, 2022) & (ZM, 2023)& (ZM, 2024)& (ZW, 2024)\\
         \hline
         Within-zone & 0.198& 0.145& 0.143 &0.261\\
         Cross-zone &0.666& 0.201& 0.404& 0.448 \\
         \hline
    \end{tabularx}
\end{table}

\section{Learning the control function}\label{sec:cross_zones}
The theoretical discussion in Section~\ref{sec:ppi} suggests learning the control function $f$ in each zone by using the field observations within the zone to predict the mean ground truth yield $Y$ from the covariates $X$ and the computer vision model predictions $\hat{Y}$.
However, we can exploit the fact that we have observations across many zones to improve the stability of the learned $f$.
While heterogeneity across zones means that pooling observations to learn $f$ likely introduces asymptotic bias in learning the true optimal control function within each zone,
it reduces the finite sample variance in the learned $f$,
translating to improved finite sample precision for the downstream PPI estimator $\hat{\theta}_{\ppipp}$.
We emphasize that this bias does not translate to asymptotic bias in $\hat{\theta}_{\ppipp}$,
but can cause it to have a higher asymptotic variance.

We find the best empirical performance when learning a single control function $f$ for each first-level administrative division in each country (states in Nigeria, provinces in Zambia and Zimbabwe).
We assign each zone to a single such administrative division (hereafter ``study region") based on plurality membership of the fields in that zone, with ties broken at random.
Appendix~\ref{app:control_function_scale} provides additional results for when a single control function $f$ is learned for all the zones in each (country, harvest year), or when a different control function is learned for each zone. 

Another important decision point for learning $f$ is the choice of learning algorithm.
We found that learning $f$ using regularized linear regression via the LASSO~\citep{tibshirani1996regression} leads to better results on our data than more complex learning algorithms such as random forests~\citep{breiman2001random} (Appendix~\ref{app:algorithm}).
We choose the LASSO penalty factor via 5-fold cross validation as implemented by the \texttt{glmnet} package in R~\citep{friedman2010regularization}. 
We include the photo model predictions $\hat{Y}_{ji}$ linearly,
and the covariates $X_{ji}$ are expanded to include interactions up to second order in the LASSO regression.

Algorithm~\ref{alg:full_pipeline} summarizes our proposed methodology from end to end.
In Theorem~\ref{thm:main} below,
we formally justify that the final PPI++ estimator from Algorithm~\ref{alg:full_pipeline} attains the same asymptotic variance as an oracle PPI++ estimator that doesn't need to estimate the optimal coefficient $\lambda$ from~\eqref{eq:lambda_hat}.
The need for this result comes from our specific choices for learning the control function $f$,
so that $f$ can be viewed neither as fully nonrandom~\citep{angelopoulos2023prediction, angelopoulos2023ppi++} nor as fully compliant with the cross-PPI paradigm~\citep{zrnic2024cross, ji2025predictions}.
We provide a full proof of this result and a brief technical discussion in Appendix~\ref{app:proof}.

\begin{theorem}
\label{thm:main}
In the setting of Algorithm~\ref{alg:full_pipeline},
suppose that for each zone $j=1,\ldots,J$,
the sample sizes $n_j$ and $N_j$ satisfy $n_j/N_j \rightarrow \rho_j \in (0,\infty)$ as $N_j \rightarrow \infty$,
the labeled observations $\{(Y_{ji},V_{ji},X_{ji})\}_{i=1}^{n_j}$ are independent and identically distributed (i.i.d.) and independent of the unlabeled observations $\{(V_{ji},X_{ji})\}_{i=n_j+1}^{n_j+N_j}$ (which are also i.i.d.),
and the photos and covariates have the same distribution between the labeled and unlabeled datasets (i.e., $\{(V_{ji},X_{ji})\}_{i=1}^{n_j+N_j}$ are i.i.d.).
Further assume that the photo models $\hat{g}^{(-k)}(\cdot), k=1,\ldots,K$ converge in mean square to a common square integrable function $g(\cdot)$,
$\e[\|X_{ji}\|_2^2] < \infty$, $\e[Y_{ji}^2] < \infty$, and $\hat{\beta}_{r(j)} \overset{p}{\to} \beta_{r(j)}$ as $N_j \rightarrow \infty$,
where $r(j) \in \{1,\ldots,R\}$ is the study region for zone $j$.
Then for $\theta_j=\e[Y_{ji}]$ we have
\[
\begin{split}
&\sqrt{n_j}(\hat{\theta}_{\ppipp,j} - \theta_j) \overset{d}{\to} \mathcal{N}(0,V_j),\\ &V_j= \rho_j \lambda_j^2 \var(f_{r(j)}^*(W_{ji})) + \var\left(Y_{ji}-\lambda_j f_{r(j)}^*(W_{ji})\right)
\end{split}
\]
where $f_{r(j)}^*(W_{ji}) = (1,g(V_{ji}),X_{ji})\beta_{r(j)}$ and $\hat{\lambda}_j \stackrel{p}{\rightarrow} \lambda_j=\frac{\cov\left(Y_{ji},f_{r(j)}^*(W_{ji})\right)}{(1+\rho_j)\var(f_{r(j)}^*(W_{ji}))}$ if $\var(f^*(W_{ji})) >0$.
\end{theorem}

Supposing that the limiting control function $f_{r(j)}^*$ defined in the statement of Theorem~\ref{thm:main} is in fact the conditional mean function $\mu_j(\cdot) = \e_j(Y \mid W=\cdot)$ (here the subscript $j$ denotes the expectation is taken over the distribution of the fields in zone $j$)
we can take the ratio of the variance $\var(Y_{ji})/n_j$ $V_j$ of $\hat{\theta}_{\lbl}$,
the sample average of the labeled field yields,
to $V_j/n_j$,
the asymptotic variance of our PPI++ estimator from
Theorem~\ref{thm:main},
to compute an asymptotic relative efficiency
in terms of
\begin{align*}
    R_j^2 = \frac{\cov(\mu_j(W_{ji}),Y_{ji})}{\var(Y_{ji})} &= \frac{\var(\mu_j(W_{ji}))}{\var(Y_{ji})}\\ &= 1-\frac{\var(Y_{ji}-\mu_j(W_{ji}))}{\var(Y_{ji})},
\end{align*}
the $R^2$ of the oracle prediction function $\mu_j$ for predicting the ground truth yields $Y$:
\begin{align}
\text{RE} & =  \frac{\var(Y_{ji})/n_j}{V_j/n_j}  \nonumber \\
& = \frac{1}{\rho_j\lambda_j^2\frac{\var(\mu_j(W_{ji}))}{\var(Y_{ji})} + \frac{\var(Y_{ji}-\lambda_j\mu_j(W_{ji}))}{\var(Y_{ji})}} \nonumber \\
& = \frac{1}{\rho_j\lambda_j^2R_j^2 + \frac{\var(Y_{ji}-\mu_j(W_{ji})) + (1-\lambda_j)^2\var(\mu_j(W_{ji}))}{\var(Y_{ji})}} \nonumber \\
& = \frac{1}{1+R_j^2 \left[\rho_j\lambda_j^2 - 1 + (1-\lambda_j)^2\right]} \nonumber \\ 
& \approx \left(1-R_j^2\frac{N_j}{N_j+n_j}\right)^{-1} \label{eq:ppipp_rel_eff}
\end{align}
where the last step follows from $\lambda_j=(1+\rho_j)^{-1}$ and $\rho_j \approx n_j/N_j$. 
Equation~\eqref{eq:ppipp_rel_eff} shows how higher signal in the photos (i.e., greater attainable zone-level $R^2$) directly improves the performance of our downstream PPI++ estimator,
justifying our use of within-zone $R^2$ as our evaluation metric in checkpointing our computer vision models,
as discussed in Section~\ref{sec:photo_prediction_model}.
We also observe from~\eqref{eq:ppipp_rel_eff} that full potential of the photos is harnessed when the number of unlabeled fields is much larger than the number of labeled fields ($N_j \gg n_j$).

\section{Results}\label{sec:results}

\begin{figure}[!t]
\centering
\includegraphics[width=\linewidth]{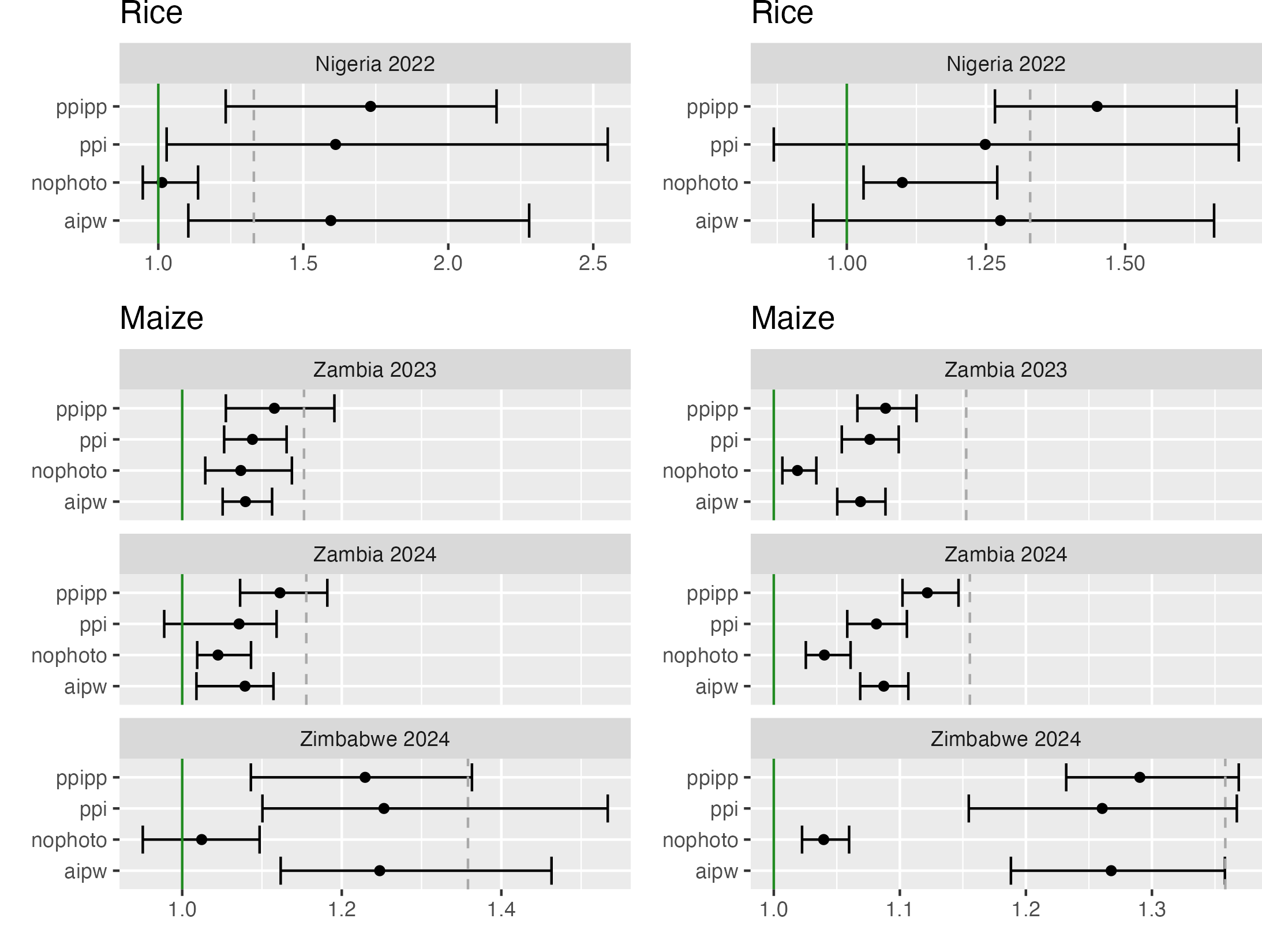}
\caption{The estimated ``MSE-based" (left) and ``CI-based" (right) relative efficiencies of the proposed method (\texttt{ppipp}) and other alternatives described in the text for each (country, harvest year) pair.
The dots are point estimates for each relative efficiency, and computed as averages of squared error ratios (left) and squared CI width ratios (right) across all synthetic bootstrap datasets for all zones.
The error bars are 95\% BCa bootstrap confidence intervals for the true relative efficiencies,
where the uncertainty stems from having only sampled a finite number of zones in each (country, harvest year) pair.
The dashed vertical grey lines indicate the theoretical asymptotic relative efficiencies of the PPI++ estimator based on~\eqref{eq:ppipp_rel_eff} with $N_j/n_j=4$ and $R_j^2$ estimated in each zone by the squared Pearson correlations between the computer vision model predictions and ground truth yields.}
\label{fig:rel_eff}
\end{figure}

We evaluate Algorithm~\ref{alg:full_pipeline} using our dataset by sampling with replacement to generate synthetic datasets respecting the observed distribution of ground truth yields and photos.
The linked code repository following the abstract reproduces these experiments.
For each zone,
we draw a random sample of $n$ fields and photos with replacement
(where $n$ is the number of fields in the zone)
to form synthetic labeled data.
By the bootstrap principle,
for evaluation purposes
the (sample) average ground truth yield of the original fields can be viewed as the unknown estimand of interest $\theta$.
We then take an independent sample of $N=4n$ photos with replacement to form the synthetic unlabeled data.
The larger amount of unlabeled data reflects the desired future setting with abundant field photos and limited crop cuts.
We repeat all of these sampling steps 10 times per zone to get more precise performance estimates.
For computational efficiency, we do not retrain the vision models across resamples,
but we do refit the control functions via the cross-validated LASSO.

We compare alternative estimators including AIPW (\texttt{aipw}),
the original PPI estimator that takes $\hat{\lambda}=1$ in~\eqref{eq:ppipp} (\texttt{ppi}),
and a variant of PPI++ that ignores the photos (\texttt{nophoto}) --- instead only adjusting for latitude and longitude  ---
to our proposed method based on the PPI++ estimator (\texttt{ppipp}).
Confidence intervals for the average yield based on each of these estimators are computed following Algorithm~\ref{alg:bootstrap} with $B=1000$,
replacing the calls to the function \texttt{PPI++} of Algorithm~\ref{alg:estimator} with calls to functions computing the relevant estimator as needed.
Our evaluation metrics for each estimator are relative efficiencies comparing with $\hat{\theta}_{\lbl}$,
the sample average of the labeled fields.
We consider both an ``MSE-based relative efficiency",
defined as the ratio of the mean squared error (MSE) of $\hat{\theta}_{\lbl}$ across all zones in a given (country, year) to the MSE of the relevant estimator,
and a ``CI-based relative efficiency,"
defined as the squared ratio of the average widths of the respective confidence intervals.
Both of these relative efficiencies can be viewed as effective sample size multipliers. 
By first-order asymptotics, they should be equivalent and independent of the zone size $n$ for sufficiently large $n$.
For the PPI++ estimator,
they can be approximated (optimistically, assuming $\hat{f}_j \approx \mu_j$ for all $j$) by averaging the right-hand side of~\eqref{eq:ppipp_rel_eff} across all zones $j$.

In Fig.~\ref{fig:rel_eff} we see that our proposed PPI++ estimator has the lowest average mean squared error (highest MSE-based relative efficiency) across all (country, harvest year) pairs except for (Zimbabwe, 2024), where the AIPW and PPI estimators edge out PPI++ very slightly.
The relative efficiencies are significantly larger than 1,
indicating strong statistical evidence for improvement over the baseline.
In particular,
the MSE-based relative efficiencies imply a
73\% increase in the effective sample size from using PPI++ to estimate average rice yields in (Nigeria, 2022),
along with 12\%, 12\%, and 23\% increases in the effective sample sizes for estimating average maize yields in (Zambia, 2023), (Zambia, 2024), and (Zimbabwe, 2024), respectively.
The markedly superior empirical performance on rice compared to maize is not consistent with the theoretical estimates based on~\eqref{eq:ppipp_rel_eff},
though these estimates all lie within the error bars for the empirical MSE-based relative efficiency estimates for the PPI++ estimator for all four (country, harvest year) pairs.
One factor that would favor empirical performance on rice is that our rice yields are noticeably less skewed than our maize yields (Fig.~\ref{fig:yields}).
It is clear that adjusting for only latitude and longitude and ignoring the photos (the \texttt{nophoto} estimator) performs worse than the proposed PPI++ estimator,
demonstrating the value of the photos and computer vision model,
although we see that $\hat{\theta}_{\nophoto}$ provides a statistically significant improvement over $\hat{\theta}_{\lbl}$ at the unadjusted 95\% confidence level in both Zambia years,
suggesting spatial coordinates have some predictive power independent of 
photos.

The PPI++ estimator also has the narrowest CI's among all methods considered for all (country, harvest year) pairs,
supporting the conclusions from the MSE-based relative efficiencies.
We see in Table~\ref{table:coverage} that this narrowness does come at some slight cost to coverage, 
though the empirical coverage differential between our intervals based on PPI++ and any of the other estimators is no more than 1\% in all of the (country, harvest year) pairs.
For rice yields in (Nigeria, 2022),
the empirical coverage for all estimators exceeds the nominal 95\% level,
while we observe slight undercoverage for all estimators for the three maize (country, harvest year) pairs,
again consistent with the skewness in the maize yields~\ref{fig:yields}.
We show in Appendix~\ref{app:ci_type_compare} that our use of BCa bootstrap confidence intervals improves the empirical coverage over both standard CLT-based intervals
and the percentile bootstrap intervals suggested by~\citet{zrnic2024note}.

\begin{figure}[!t]
\centering
\includegraphics[width=\linewidth]{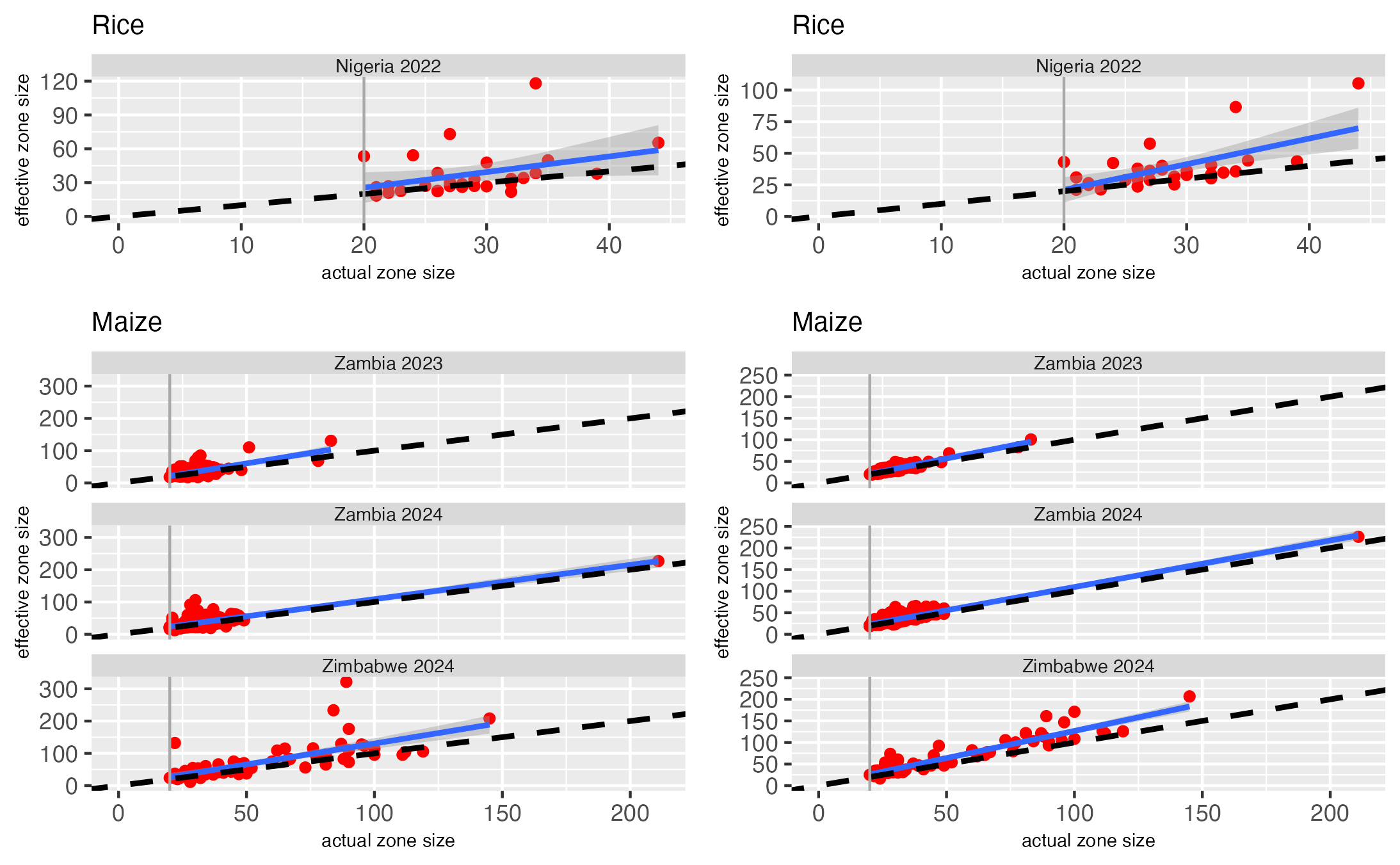}
\caption{The estimated effective sample size
is plotted on each vertical axis against the actual zone size on the horizontal axis for each (country, harvest year) pair.
Effective sample size is defined for each zone as the ratio of squared error (left) or squared ratio of CI width (right) from $\hat{\theta}_{\lbl}$ to that from $\hat{\theta}_{\ppipp}$, multiplied by the zone size.
The blue lines are ordinary linear regression fits to the scatterplots and the dashed lines pass through the origin with slope 1.}
\label{fig:zone_level_plots}
\end{figure}

Continuing the interpretation of our relative efficiency metrics as sample size multipliers,
Fig.~\ref{fig:zone_level_plots} suggests that the PPI++ estimator improves the effective sample size --- according to both MSE and CI width --- for all zone sizes and all (country, harvest year) pairs.
Notably,
the improvement expected in large samples from Theorem~\ref{thm:main} is observed empirically in zones as small as our \emph{a priori} cutoff of 20 fields.

\begin{table}[!t]
\centering
\caption{The empirical coverage probabilities of the various nominal 95\% confidence intervals across zones in each (country, harvest year).
}
\label{table:coverage}
\begin{tabularx}{\linewidth}{|c|X|X|X|X|}
\hline
Estimator & (NG, 2022) & (ZM, 2023) & (ZM, 2024) & (ZW, 2024) \\
\hline
$\hat{\theta}_{\lbl}$ & 0.962 & 0.939 & 0.929 & 0.923 \\
$\hat{\theta}_{\aipw}$ & 0.959 & 0.936 & 0.923 & 0.926 \\
$\hat{\theta}_{\ppi}$ & 0.955 & 0.940 & 0.928 & 0.922 \\
$\hat{\theta}_{\ppipp}$ & 0.952 & 0.931 & 0.927 & 0.916 \\
$\hat{\theta}_{\nophoto}$ & 0.952 & 0.939 & 0.931 & 0.921 \\
\hline
\end{tabularx}
\end{table}
\section{Summary and discussion}\label{sec:discussion}

We have shown that PPI provides a mathematically principled and empirically performant way to use a computer vision model to leverage easily obtained field photos to supplement limited ground-truth field observations for precisely estimating average yields.
The method provably does not introduce any asymptotic bias in the point estimates,
nor can it decrease the asymptotic precision relative to taking the simple average of ground truth observations.
Key implementation decisions such as the supervised learning algorithm for fitting the control function,
the scale at which observations are pooled to train the control function,
and the method used to compute confidence intervals can have practically relevant impacts on performance.
Thus, we recommend that prior to deployment,
a user of our method validates their choices
using simulations that reflect their data-generating process,
as we have in Section~\ref{sec:results}.

Standardized photo-taking guidelines were absent at the time the photos were taken. 
The result is that photos vary significantly in terms of the portion of photo covered by the field and the zoom level of the field (see Fig.~\ref{fig:sample_photos}). 
We thus expect that a more standardized photo-taking procedure would improve performance and perhaps inform a different set of implementation choices.

The PPI framework extends to a wide range of estimands beyond average yields,
so our methods could be straightforwardly extended to estimate other quantities of interest like specific yield quantiles.
When available,
one could also integrate other sources of low-cost data besides field photos,
such as satellite imagery.

\onecolumn 

\bibliography{pula}
\pagebreak

\renewcommand{\appendixpagename}{Appendix}
\appendix
\appendixpage

\section{Equivalence of RePPI and PPI++ for estimating the mean}\label{app:reppi}
Recently,~\citet{ji2025predictions} proposed a ``recalibrated PPI (RePPI)" procedure that asymptotically dominates the PPI++ estimator,
as supported by a ``discrete predictions" example in their Section 4.3.
However, we show this improvement would only exist in our setting if we were comparing to a different PPI++ estimator than ours
that used the photo model predictions $\hat{g}(V_i)$ directly as the control function $f$ as opposed to additionally including our LASSO-based step of learning $f$ using both the photo model predictions and the additional covariates $X_i$.
It turns out that this latter step is equivalent to the recalibration used in RePPI,
as we explore below.

Recalibrated PPI estimators are defined for any estimand $\theta$ whose true value $\theta^*$ minimizes a suitably smooth expected loss:
\begin{equation}
\label{eq:M_estimand}
\theta^* = \argmin_{\theta \in \Theta} \e[\ell_{\theta}(X,Y)]
\end{equation}
(cf. equation (3) of~\citet{ji2025predictions}).
This framework assumes access to covariates $X$ and
machine learning predictions $\hat{Y}$ 
in both the labeled and unlabeled datasets,
and true outcomes $Y$ in only the labeled dataset.
The predictions $\hat{Y}$ would correspond to the photo model predictions $\hat{g}(V_i)$ in our setting,
following the notation in the main text.
For our estimand $\theta=\e[Y]$,
the true value $\theta^*$ satisfies~\eqref{eq:M_estimand} with $\Theta=\real$ and $\ell_{\theta}(X,Y)=(1/2)(Y-\theta)^2$.
Then
by Theorem 1 of~\citet{ji2025predictions} an optimal RePPI estimator takes the form
\begin{equation}
\label{eq:reppi}
\hat{\theta}_{\reppi} = \argmin_{\theta} \frac{1}{n} \sum_{i=1}^n \ell_{\theta}(X_i,Y_i)  - \bigg(\frac{1}{n} \sum_{i=1}^n g_{\theta}(X_i;\hat{Y}_i)-\frac{1}{N}\sum_{i=n+1}^{n+N} g_{\theta}(X_i;\hat{Y}_i)\bigg)    
\end{equation}
for
\begin{align*}
g_{\theta}(X,\hat{Y}) & = \frac{N}{n+N}\theta\e[\nabla \ell_{\theta}^*(X,Y) \mid X,\hat{Y}] = \frac{N}{n+N}\theta(\theta^*-\mu(X,\hat{Y})), \\
\mu(x,\hat{y}) & = \e[Y \mid X=x,Y=\hat{y}].
\end{align*}
The function $g_{\theta}$ has two unknown components: $\theta^*$ and the function $\mu$.
The $\theta^*$ term ultimately cancels out in the RePPI objective~\eqref{eq:reppi},
so it remains to estimate $\mu$. 
~\citet{ji2025predictions} propose doing so via a ``flexible machine learning method,"
i.e., fitting a nonparametric regression to predict $Y$ from $X$ and $\hat{Y}$.
Mean square convergence of this regression at any rate suffices to ensure the resulting RePPI estimator remains optimal.
This regression is precisely the principle behind how we learn the control function $f$ for our PPI++ estimator
(for our data, we used the LASSO rather than nonparametric regression for the empirical and small sample reasons discussed in Section~\ref{sec:cross_zones} and validated in Appendix~\ref{app:algorithm}).

Once the function $\mu$ is estimated,
we obtain an estimate of the function $g_{\theta}$,
which RePPI then pre-multiplies by an estimated ``optimal" coefficient $\hat{M}$
before plugging into~\eqref{eq:reppi}.
This coefficient $\hat{M}$ can be seen to be identical to $\hat{\lambda}$ from~\eqref{eq:lambda_hat},
and then solving the minimization~\eqref{eq:reppi} precisely yields our PPI++ estimator~\eqref{eq:ppipp}.

\section{Additional results}
We collect some additional results on the performance of our proposed methods with different implementation choices,
justifying the decisions made in the main text.
\subsection{Confidence interval methodology}
\label{app:ci_type_compare}
\begin{figure}[!t]
\centering
\includegraphics[width=0.75\linewidth]{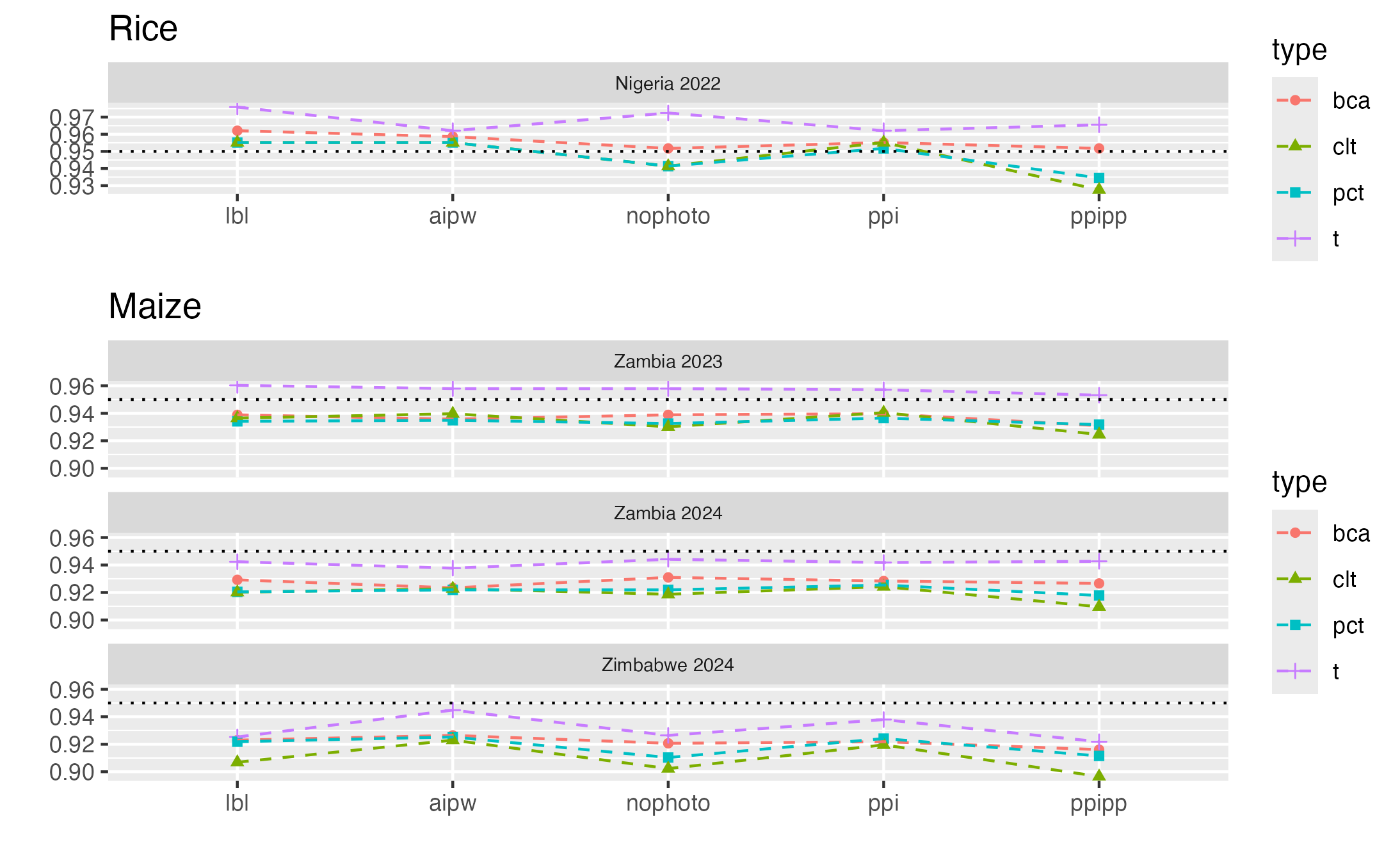} \\
\caption{The empirical coverage probabilities of the four different nominal 95\% confidence intervals based around each estimator described in Appendix~\ref{app:ci_type_compare}, broken down by (country, harvest year)
}
\label{fig:ci_type_coverage_compare}
\end{figure}

\begin{figure}[!tb]
\centering
\includegraphics[width=0.75\linewidth]{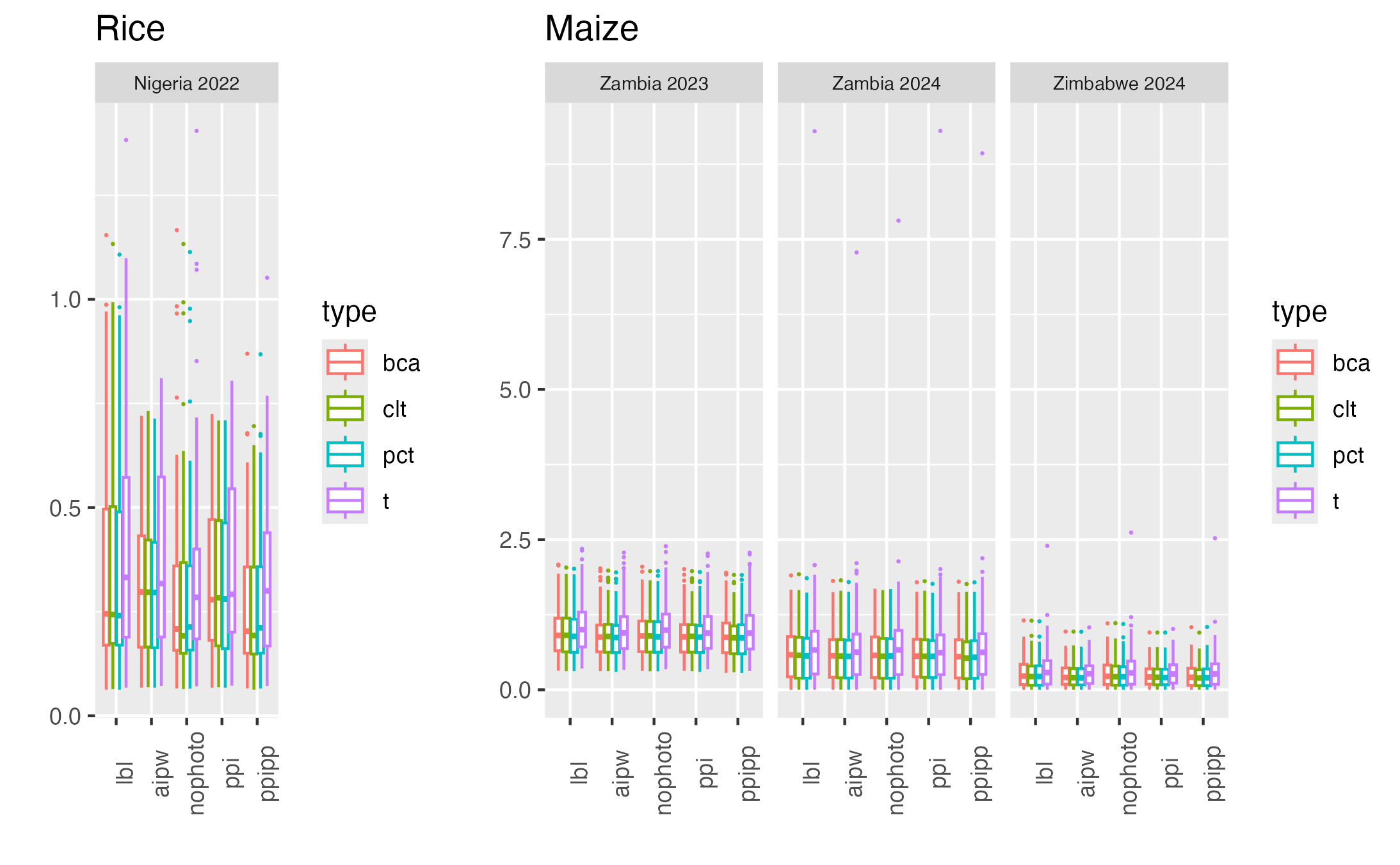} \\
\caption{The distribution of the lengths of the confidence intervals in Fig.~\ref{fig:ci_type_coverage_compare}, in $\texttt{mt ha}^{-1}$
}
\label{fig:ci_type_width_compare}
\end{figure}

Figures~\ref{fig:ci_type_coverage_compare} and~\ref{fig:ci_type_width_compare} examine the empirical coverage probabilities and widths,
respectively,
of four different nominal 95\% confidence intervals computed for each estimator.
One of these methods is the BCa bootstrap from Algorithm~\ref{alg:bootstrap},
following the main text.
The other three methods all use the same point estimators
but use non-bootstrap intervals based on the normal approximation (\texttt{clt}) or the bootstrap-$t$ (\texttt{t}) or the percentile bootstrap (\texttt{pct}) in lieu of the BCa bootstrap.

Compared to the BCa bootstrap intervals,
the CLT intervals tend to be of similar length but suffer from a markedly lower coverage rate for $\hat{\theta}_{\lbl}$, $\hat{\theta}_{\nophoto}$, and $\hat{\theta}_{\ppipp}$ across all (country, harvest year) pairs.
On the other hand, the CLT-based intervals for the AIPW and PPI estimators seem to cover about as well as (indeed, slightly more than) the BCa intervals, while being slightly longer.
This suggests that the CLT-based intervals may be less robust to the finite-sample error in estimating the coefficient $\hat{\lambda}$,
as required by the PPI++ estimators.
We ultimately chose the BCa bootstrap intervals over the CLT-based intervals for their superior theoretical properties (i.e., second-order asymptotic coverage) and better empirical performance for the $\hat{\theta}_{\lbl}$ and $\hat{\theta}_{\ppipp}$ estimators,
the two estimators of most interest.
The percentile bootstrap intervals are also of very similar lengths as the BCa intervals but generally exhibit slightly worse coverage properties.
The bootstrap-$t$ intervals have the best empirical coverage properties for all methods and estimators,
but occasionally produce unusuably long intervals,
most notably in (Zambia, 2024) and (Zimbabwe, 2024)
(recall that the ground truth yields for all fields were between 0 and 10.7 \texttt{mt $\text{ha}^{-1}$}).
This phenomenon was noted previously in the simulations of~\citet{diciccio1996bootstrap} who advocate for the use of BCa intervals over the bootstrap-$t$ in practice for this reason,
noting that both methods share the same theoretical second-order asymptotic coverage guarantees.

\subsection{Pooling across zones for learning control functions}
\label{app:control_function_scale}

\begin{figure}[!t]
\centering
\includegraphics[width=\linewidth]{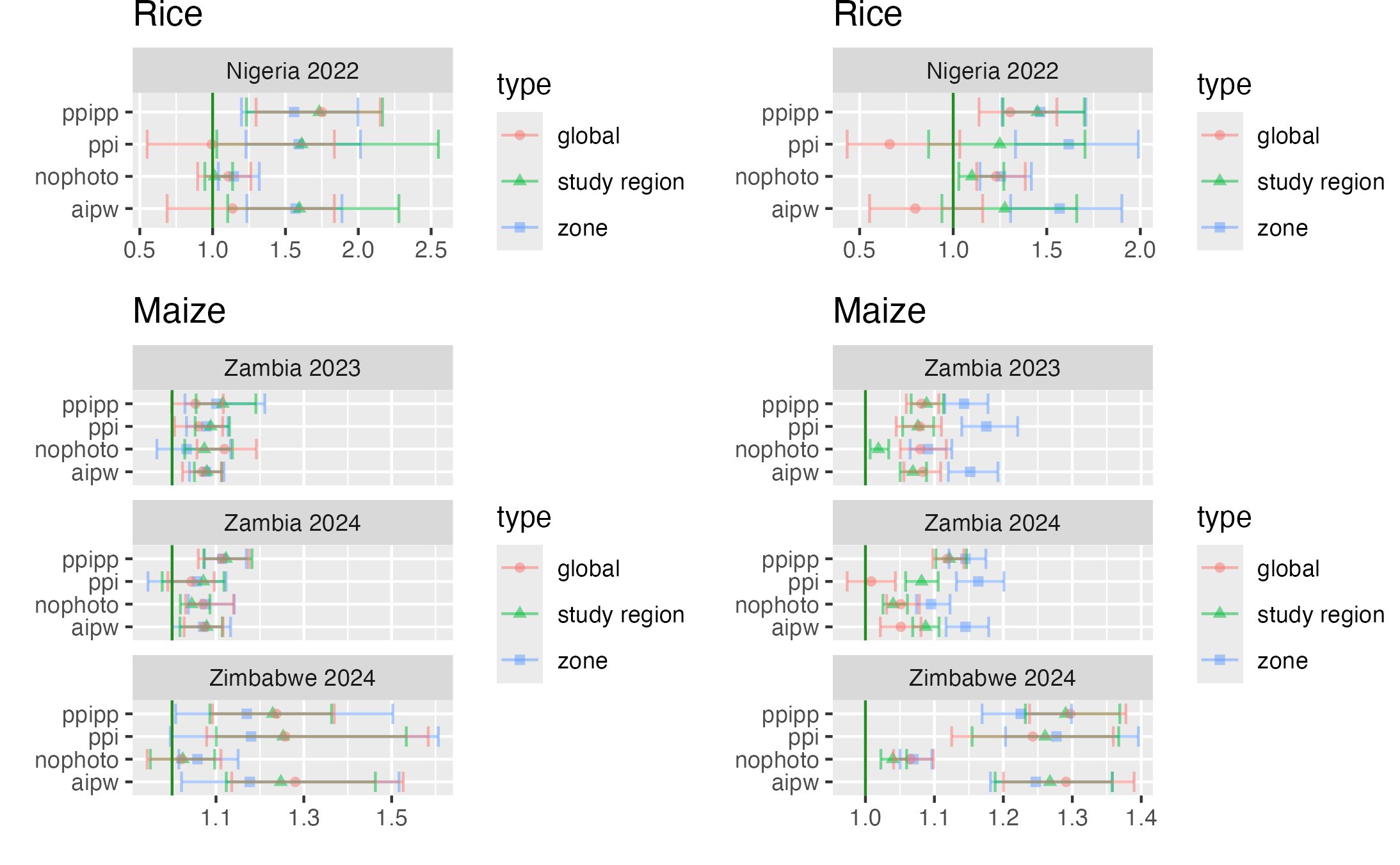} \\
\caption{Estimated MSE-based (left) and CI-based (right) relative efficiencies,
as defined in Section~\ref{sec:results},
for control functions learned at three different scales: global, study region, and zone.
The results for the study region scale are identical to those in Fig.~\ref{fig:rel_eff} in the main text.
}
\label{fig:rel_eff_scale_compare}
\end{figure}

Here we present additional results showing the effect of varying the scale at which control functions are learned --- that is,
the degree to which observations are pooled across zones for learning the control function for our PPI estimators.
We consider pooling observations across each study region (corresponding to the results in the main text),
as well as learning a separate control function for each zone (the ``zone" scale)
and learning a single control function pooling all observations across all zones in an entire (country, harvest year) pair (the ``global" scale).

\begin{figure}[!t]
\centering
\includegraphics[width=0.75\linewidth]{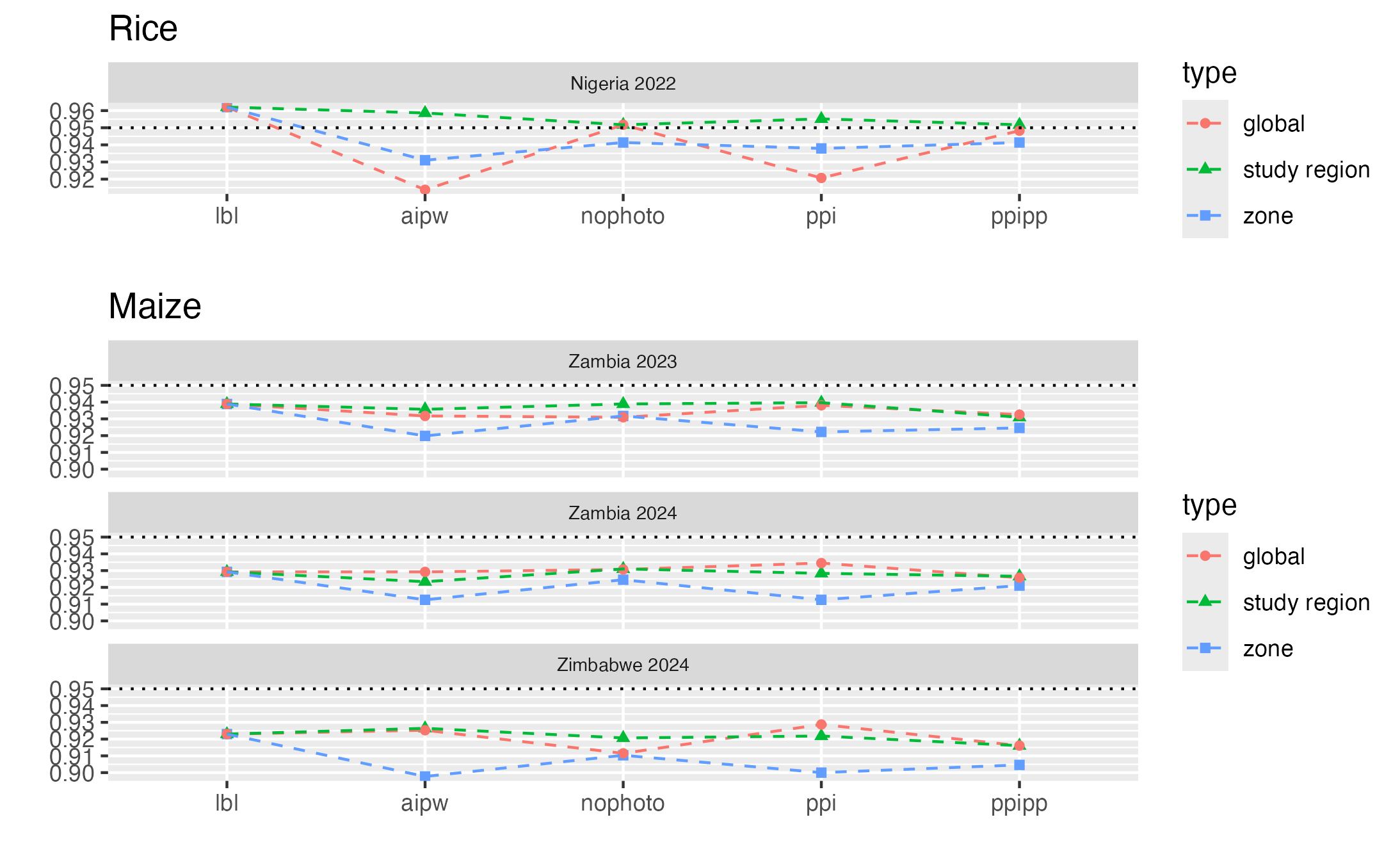} \\
\caption{
The empirical coverage probabilities of the nominal 95\% confidence intervals from the simulations in Fig.~\ref{fig:rel_eff_scale_compare}
}
\label{fig:rel_eff_scale_coverage}
\end{figure} 

Fig.~\ref{fig:rel_eff_scale_compare} shows that for each estimator,
the zone scale point estimates perform worse than the study region scale estimates,
sometimes substantially so.
On the other hand,
the zone scale estimates come with narrower confidence intervals (except in Zimbabwe, 2024);
however,
this comes at substantial cost to coverage (Fig.~\ref{fig:rel_eff_scale_coverage}).
We would expect this theoretically,
since the asymptotic coverage guarantees of the confidence intervals are based on the fact that the variability in the learned control function is first-order negligible.
This asymptotic approximation would of course be most problematic when the training sample for the control function is relatively small,
i.e., on the zone scale.

The global scale
performs mostly similarly to the study region scale in terms of both relative efficiency and confidence interval coverage for the PPI++ estimator.
We ended up choosing the study region scale for noticeably better point estimate performance in (Zambia, 2023)
and somewhat more consistent coverage properties across all estimators considered (Fig.~\ref{fig:rel_eff_scale_coverage}).
Overall,
the results demonstrate the finite sample benefits of pooling information across multiple zones for learning the control function.
Even if pooling means the asymptotic limit of the learned control function is not optimal 
(in terms of minimizing the asymptotic variance of the finial PPI++ estimator)
for any given zone,
the increased finite sample stability of the learned control functions
more than compensates for this and improves the empirical final performance of the PPI++ point estimates.
If the zones were larger,
we would expect this trade-off to reverse,
with the zone scale performing relatively better than the study region and global scales.

\subsection{Learning algorithm for fitting control functions}
\label{app:algorithm}

\begin{figure}[!t]
\centering
\includegraphics[width=\linewidth]{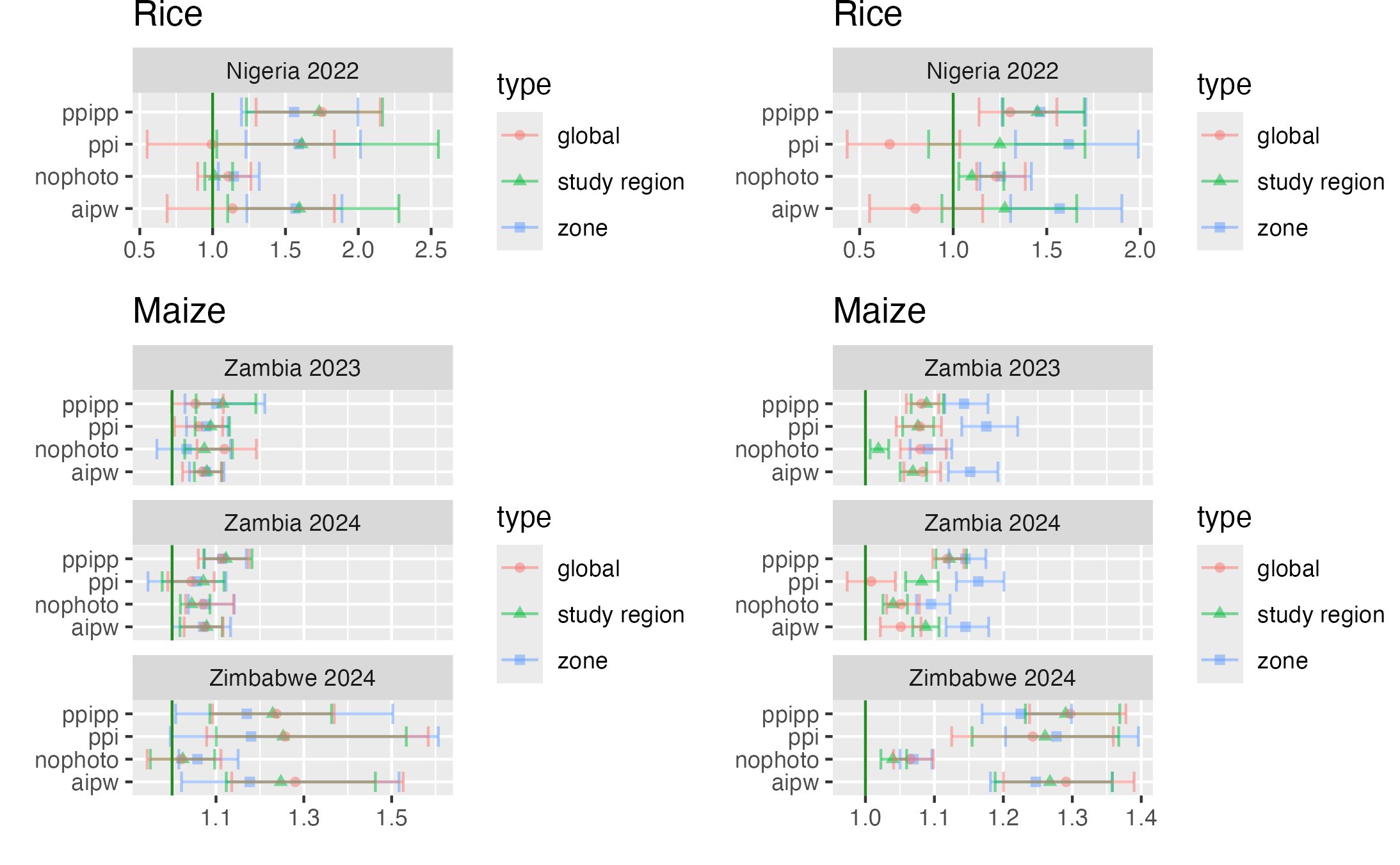} \\
\caption{Estimated MSE-based (left) and CI-based (right) relative efficiencies,
as defined in Section~\ref{sec:results},
for control functions learned using the cross-validated LASSO as described in the main text (\texttt{glmnet}) and for control functions learned using random forests (\texttt{rf}),
as described in Appendix~\ref{app:algorithm}.
The results for \texttt{glmnet} are identical to those in Fig.~\ref{fig:rel_eff} in the main text.
}
\label{fig:rel_eff_alg_compare}
\end{figure}

In Fig.~\ref{fig:rel_eff_alg_compare} we see that learning the control function $f$ using a random forest (as implemented by the \texttt{regression\_forest} function in the \texttt{grf} package in R~\citep{tibshirani2018package}) leads to noticeably less precise point estimates and wider confidence intervals than using the cross-validated LASSO advocated in the main text.
The coverage also seems to be generally worse.
While the random forest enables more flexible control functions to be learned,
this comes at the cost of finite sample variance,
which appears to dominate on our data.
With larger study regions and/or less noisy yields, 
we might expect these more flexible algorithms to perform better.

\begin{figure}[!t]
\centering
\includegraphics[width=\linewidth]{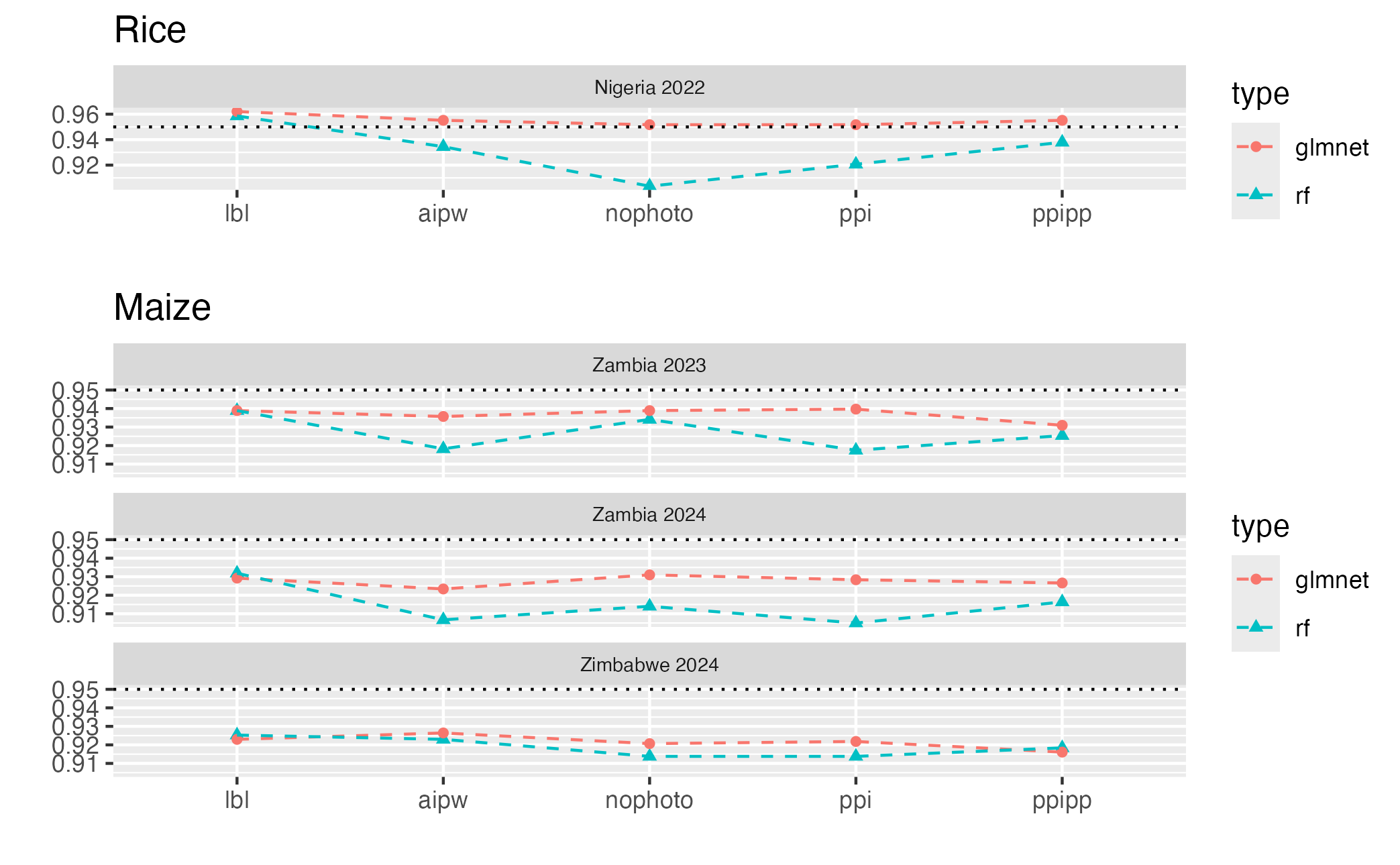} \\
\caption{
The empirical coverage probabilities of the nominal 95\% confidence intervals from the simulations in Fig.~\ref{fig:rel_eff_alg_compare}
}
\label{fig:rel_eff_alg_coverage}
\end{figure} 

\section{Discussion and proof of Theorem~\ref{thm:main}}\label{app:proof}
As alluded to in the main text,
the need for Theorem~\ref{thm:main} comes from the fact that we use cross-fitting when training the vision models but not to learn the final control function from the model predictions and spatial covariates,
so that we are neither in the cross-PPI setting nor the ``standard" PPI setting studied theoretically by previous authors.
The motivation for this hybrid cross-fitting is as follows.
As neural networks,
the vision models can have fairly high function complexity,
motivating us to use cross-fitting to control the error in the final estimator coming from the finite-sample variance in the trained vision models $\hat{g}^{(-k)}$,
just as in cross-PPI and double machine learning\mbox{~\citep{chernozhukov2018double}}.
On the other hand,
the final control function is fit using a low-dimensional linear model,
and hence of sufficiently low complexity that the propagation of the finite sample error in the learned control function into our final estimator $\hat{\theta}_{\ppipp}$ can be controlled without cross-fitting.
While introducing an additional layer of cross-fitting for this step would still enable valid theoretical results,
it adds algorithmic and computational complexity.
Additionally,
several works
have suggested that cross-fitting degrades empirical performance in small to moderate sample sizes\mbox{~\citep{chen2022debiased,okasa2022meta}}.

If we believe the various i.i.d. assumptions,
the conditions of Theorem~\ref{thm:main} are quite mild and likely to be satisfied in practice.
Note that we do not impose any rate requirements on how quickly the photo model predictions $\hat{g}^{(-k)}$ or the LASSO coefficient $\hat{\beta}$ need to converge.
We can get away with arbitrarily slow rates because the assumption that the distribution of $W_{ji}=(V_{ji},X_{ji})$ is invariant between the labeled and unlabeled datasets is equivalent to knowing the labeling propensity score (the probability that a randomly selected observation,
pooling together the labeled and unlabeled observations,
is a labeled observation conditional on $W=w$)
is constant in $w$ (namely $n/(n+N)$).
This phenomenon is well-known for the AIPW estimator,
where various authors have shown that when propensity scores are known,
any rate of convergence in estimating the outcome mean functions suffices for the estimator to be asymptotically equivalent to an oracle that knows the functions~\citep{rothe_value_2016,chernozhukov2018double}.
One simple way to ensure theoretically that $\hat{\beta}$ converges in a low-dimensional setting ($n_j \gg p$),
given that the $\hat{g}^{(-k)}$ converge,
is to cap the penalty parameter in the LASSO regression by constants converging to 0 (at any arbitrarily slow rate) with the sample sizes $n_j,N_j$.
In a high-dimensional setting,
we could modify the cross-validated LASSO algorithm to the highly adaptive LASSO which comes with robust convergence guarantees in the space of c\`adl\`ag functions~\citep{benkeser2016highly,van2017generally}.

\begin{proof}
We fix an arbitrary zone $j$ and for brevity omit the $r(j)$ subscripts in our notation,
(e.g., writing $\hat{\beta} \equiv \hat{\beta}_{r(j)}$, $\beta \equiv \beta_{r(j)}$, $f \equiv f_{r(j)}$, and $f^*_{r(j)} \equiv f^*$).
Our proof of Theorem~\ref{thm:main} proceeds by defining an oracle variant $\hat{\theta}_{\ppipp,j}^*$ of our estimator $\hat{\theta}_{\ppipp,j}$ that plugs in the limiting control function $f^*$ and limiting coefficient $\lambda_j$:
\begin{equation}
\label{eq:oracle_ppipp}
\hat{\theta}_{\ppipp,j}^* = \frac{1}{n_j} \sum_{i=1}^{n_j} Y_{ji}- \lambda_j \left( \frac{1}{n_j} \sum_{i=1}^{n_j} f^*(W_{ji}) - \frac{1}{N_j} \sum_{i=n_j+1}^{n_j+N_j} f^*(W_{ji}) \right)
\end{equation}
We will show the oracle satisfies the central limit theorem of Theorem~\ref{thm:main} and that the difference between the oracle $\hat{\theta}_{\ppipp,j}^*$ and the actual estimator $\hat{\theta}_{\ppipp,j}$ is asymptotically negligible (i.e., it converges to zero in probability at a rate faster than $n_j^{-1/2}$).
In the course of showing the latter we will also prove that $\hat{\lambda}_j \stackrel{p}{\rightarrow} \lambda_j$.

We write
\[
    \hat{\theta}_{\ppipp,j}^* = \frac{1}{n_j} \sum_{i=1}^{n_j} (Y_{ji}-\lambda_jf^*(W_{ji})) + \frac{1}{N_j} \sum_{i=n_j+1}^{n_j+N_j} \lambda_jf^*(W_{ji})
\]
With $\e\left[f^*(W_{ji})f^*(W_{ji})^{\top}\right]$ existing by the assumptions that $g(V_{ji})$ and $X_{ji}$ are square integrable, 
for $\tau_j=\lambda_j\e[f^*(W_{ji})]$ we have by the central limit theorem that
\[
\sqrt{n_j}\left(
\begin{bmatrix}
\frac{1}{n_j} \sum_{i=1}^{n_j} Y_{ji}-\lambda_j f^*(W_{ji}) \\
\frac{1}{N_j} \sum_{i=n_j+1}^{n_j+N_j} \lambda_j f^*(W_{ji})
\end{bmatrix}
-
\begin{bmatrix}
\theta_j-\tau_j \\
\tau_j
\end{bmatrix}
\right)
\stackrel{d}{\rightarrow} \mathcal{N}\left(
\begin{bmatrix}
0 \\
0
\end{bmatrix},
\begin{bmatrix}
\var_{\lbl}(Y_{ji} - \lambda_jf^*(W_{ji})) & 0 \\
0 & \rho_j\lambda_j^2 \var(f^*(W_{ji}))
\end{bmatrix}
\right)
\]
The oracle central limit theorem
\[
\sqrt{n_j}(\hat{\theta}_{\ppipp,j}^* - \theta_j) \stackrel{d}{\rightarrow} \mathcal{N}(0, V_j)
\]
then follows by the Delta method.

Now we write
\begin{align}
\hat{\theta}_{\ppipp,j} - \hat{\theta}_{\ppipp,j}^* & = \lambda_j\bigg(\frac{1}{n_j} \sum_{i=1}^{n_j} [f^*(W_{ji})-f(W_{ji})]  -\frac{1}{N_j} \sum_{i=n_j+1}^{n_j+N_j}[f^*(W_{ji})-f(W_{ji})]\bigg)  \label{eq:asymp_1} \\
& \quad + (\lambda_j-\hat{\lambda}_j)\bigg(\frac{1}{n_j} \sum_{i=1}^{n_j} f(W_{ji})  - \frac{1}{N_j} \sum_{i=n_j+1}^{n_j+N_j} f(W_{ji})\bigg) \label{eq:asymp_2}
\end{align}
and show each of these two terms is asymptotically negligible.
For the first term~\eqref{eq:asymp_1},
we partition $\beta=(\beta_0,\beta_1,\beta_2^{\top})^{\top}$ where $\beta_0\in\real$ and $\beta_1 \in \real$ to conform with the components of $\psi(\cdot)$ and write
\begin{align}
f^*(W_{ji})-f(W_{ji}) & = \hat{\beta}_1\left(g(V_{ji})-\hat{g}^{(-k(j,i))}(V_{ji})\right)  - (\hat{\beta}-\beta)^{\top} \psi^*(W_{ji}) , \quad i \leq n_j \label{eq:f_diff_lbl} \\
f^*(W_{ji})-f(W_{ji}) & = \hat{\beta}_1\left(g(V_{ji})-\frac{1}{K} \sum_{k=1}^K \hat{g}^{(-k)}(V_{ji})\right) - (\hat{\beta}-\beta)^{\top}\psi^*(W_{ji}) , \quad i > n_j \label{eq:f_diff_unlbl}
\end{align}
where $k(j,i) \in \{1,\ldots,K\}$ is the fold to which $(j,i)$ belongs and $\psi^*(W_{ji}) = (1,g(V_{ji}),X_{ji}^{\top})^{\top}$.
Thus
\begin{align*}
\frac{1}{n_j} \sum_{i=1}^{n_j} \left[f^*(W_{ji}) - f(W_{ji})\right] - \frac{1}{N_j} \sum_{i=n_j+1}^{n_j+N_j} \left[f^*(W_{ji}) - f(W_{ji})\right]
&= \sum_{k=1}^K \frac{1}{n_j} \sum_{i : (j,i) \in \mathcal{I}_k}
    \left[\hat{\beta}_1 \left(g(V_{ji}) - \hat{g}^{(-k)}(V_{ji})\right)\right] \\
&\quad - (\hat{\beta} - \beta)^\top \frac{1}{n_j} \sum_{i=1}^{n_j} \psi^*(W_{ji}) \\
&\quad - \sum_{k=1}^K \frac{1}{K N_j} \sum_{i=n_j+1}^{n_j+N_j}
    \left[\hat{\beta}_1 \left(g(V_{ji}) - \hat{g}^{(-k)}(V_{ji})\right)\right] \\
&\quad + (\hat{\beta} - \beta)^\top \frac{1}{N_j} \sum_{i=n_j+1}^{n_j+N_j} \psi^*(W_{ji}).
\end{align*}

The second and fourth terms can be written as
\[
(\hat{\beta}-\beta)^{\top}
\begin{bmatrix}
0 \\
\frac{1}{N_j} \sum_{i=n_j+1}^{n_j+N_j} g(V_{ji}) - \frac{1}{n_j} \sum_{i=1}^{n_j} g(V_{ji}) \\
\frac{1}{N_j} \sum_{i=n_j+1}^{n_j+N_j} X_{ji} - \frac{1}{n_j} \sum_{i=1}^{n_j} X_{ji}
\end{bmatrix}
= o_p(n_j^{-1/2})
\]
where the asymptotic negligibility comes from the fact that $\hat{\beta}-\beta=o_p(1)$ by assumption and
the differences
\[
\frac{1}{N_j} \sum_{i=n_j+1}^{n_j+N_j} g(V_{ji}) - \frac{1}{n_j} \sum_{i=1}^{n_j} g(V_{ji}) \quad \text{ and } \quad \frac{1}{N_j} \sum_{i=n_j+1}^{n_j+N_j} X_{ji} - \frac{1}{n_j} \sum_{i=1}^{n_j} X_{ji}
\]
are of order $n_j^{-1/2}$,
since they both have zero mean and (co)variances on the order of $n_j^{-1}$ in view of $\{W_{ji}\}_{i=1}^{n_j+N_j}$ being i.i.d.
The remaining terms are
\[
    \sum_{k=1}^K \hat{\beta}_1 \bigg[\frac{1}{n_j} \sum_{i:(j,i) \in \ci_k} \left(g(V_{ji})-\hat{g}^{(-k)}(V_{ji})\right) - \frac{1}{KN_j} \sum_{i=n_j+1}^{n_j+N_j} \left(g(V_{ji})-\hat{g}^{(-k)}(V_{ji})\right)\bigg]
\]
and so to complete the proof that the terms~\eqref{eq:asymp_1} are $o_p(n_j^{-1/2})$ it suffices to show that
\[
    A_k = \frac{1}{n_j} \sum_{i:(j,i) \in \ci_k} \left(g(V_{ji})-\hat{g}^{(-k)}(V_{ji})\right) - \frac{1}{KN_j} \sum_{i=n_j+1}^{n_j+N_j} \left(g(V_{ji})-\hat{g}^{(-k)}(V_{ji})\right) = o_p(n_j^{-1/2})
\]
for each $k=1,\ldots,K$.
We fix any such $k$,
let $n_j^{(k)}$ be the number of indices among $(j,1),\ldots,(j,n_j)$ in fold $k$,
and note that $\hat{g}^{(-k)}$ is nonrandom conditional on $\cf^{(-k)}$,
the $\sigma$-algebra $\cf^{(-k)}$ generated by the labeled observations outside fold $k$,
i.e., $\{(Y_{ji},V_{ji},X_{ji}): (j,i) \notin \ci_k\}$.
Then
\[
\tilde{A}_k = \frac{1}{Kn_j^{(k)}}  \sum_{i:(j,i) \in \ci_k} \left(g(V_{ji})-\hat{g}^{(-k)}(V_{ji})\right) - \frac{1}{KN_j} \sum_{i=n_j+1}^{n_j+N_j} \left(g(V_{ji})-\hat{g}^{(-k)}(V_{ji})\right)
\]
satisfies $\e[\tilde{A}_k \mid \cf^{(-k)}]=0$,
while
\[
\var(\tilde{A}_k \mid \cf^{(-k)}) = \left(\frac{1}{K^2n_j^{(k)}} + \frac{1}{K^2N_j}\right)V^{(-k)}
\]
where 
\begin{align}
V^{(-k)} = \var\left(g(V_{ji})-\hat{g}^{(-k)}(V_{ji}) \mid \cf^{(-k)}\right) \nonumber \leq \e\left[\left(g(V_{ji})-\hat{g}^{(-k)}(V_{ji})\right)^2 \Big| \cf^{(-k)} \right] \nonumber = o_p(1) \label{eq:V_neg_k}
\end{align}
since $\hat{g}^{(-k)}$ converges to $g$ in mean square.
It follows that 
\[
\e[\tilde{A}_k^2 \mid \cf^{(-k)}] = \var(\tilde{A}_k \mid \cf^{(-k)}) = o_p(n_j^{-1})
\]
and so $\tilde{A}_k = o_p(n_j^{-1/2})$ by e.g., Lemma 2 of~\citet{li2024efficient}.
Additionally we claim that 
\[
    A_k-\tilde{A}_k = \left(\frac{1}{n_j}-\frac{1}{Kn_j^{(k)}}\right) \sum_{(j,i) \in \ci_k} g(V_{ji})-\hat{g}^{(-k)}(V_{ji}) = o_p(n_j^{-1}).
\]
To see this, note that $|n_j-Kn_j^{(k)}| \leq 1$ by the fact that the folds are divided as evenly as possible.
Thus
\[
n_j^{(k)}\left(\frac{1}{n_j}-\frac{1}{Kn_j^{(k)}}\right)= \frac{n_j-Kn_j^{(k)}}{Kn_j} = O_p(n_j^{-1})
\]
while 
\[
\frac{1}{n_j^{(k)}} \sum_{i:(j,i) \in \ci_k} g(V_{ji})-\hat{g}^{(-k)}(V_{ji}) = o_p(1)
\]
since
\begin{equation}
\label{eq:g_hat_vs_g}
\Bigg| \frac{1}{n_j^{(k)}} \sum_{i:(j,i) \in \ci_k} g(V_{ji})-\hat{g}^{(-k)}(V_{ji}) \Bigg| \leq \sqrt{\frac{1}{n_j^{(k)}}  \sum_{i:(j,i) \in \ci_k} \left(g(V_{ji})-\hat{g}^{(-k)}(V_{ji})\right)^2} = o_p(1)
\end{equation}
by the fact that $\hat{g}^{(-k)}$ converges to $g$ in mean square and Lemma 4 of~\citet{li2024efficient}.
So we can conclude $A_k = \tilde{A}_k + (A_k-\tilde{A}_k) = o_p(n_j^{-1/2})$,
completing the argument that
\begin{align}
\label{eq:asymp_1_redux}
\frac{1}{n_j} \sum_{i=1}^{n_j} [f^*(W_{ji})-f(W_{ji})] - \frac{1}{N_j} \sum_{i=n_j+1}^{n_j+N_j}[f^*(W_{ji})-f(W_{ji})] = o_p(n_j^{-1/2}).
\end{align}

It remains to show the terms~\eqref{eq:asymp_2} are asymptotically negligible as well.
If we assume $\hat{\lambda}_j-\lambda_j=o_p(1)$,
the terms~\eqref{eq:asymp_2} are asymptotically equivalent to
\[
(\lambda_j-\hat{\lambda}_j)\left(\frac{1}{n_j} \sum_{i=1}^{n_j} f^*(W_{ji}) - \frac{1}{N_j} \sum_{i=n_j+1}^{n_j+N_j} f^*(W_{ji})\right)
\]
by~\eqref{eq:asymp_1_redux}.
But 
\[
\frac{1}{n_j} \sum_{i=1}^{n_j} f^*(W_{ji}) - \frac{1}{N_j} \sum_{i=n_j+1}^{n_j+N_j} f^*(W_{ji}) = O_p(n_j^{-1/2})
\]
since it has mean 0 and variance
\[
\left(n_j^{-1} + N_j^{-1}\right)\var(f^*(W)) = O(n_j^{-1}).
\]
Thus to complete our proof it suffices to show that indeed, $\hat{\lambda}_j \stackrel{p}{\rightarrow} \lambda_j$.
To that end we have
\[
\hat{\lambda}_j = \frac{N_j}{n_j+N_j}\frac{\widehat{\cov}_j(Y,f(W))}{\widehat{\var}_j(f(W))}
\]
where (cf. Algorithm~\ref{alg:estimator})
\[
\widehat{\cov}_j(Y,f(W)) = \frac{1}{n_j-1} \sum_{i=1}^{n_j} (Y_{ji}-\bar{Y}_{j\cdot})(f(W_{ji})-\bar{f}_{n,j})
\]
for
\[
\bar{Y}_{j\cdot} = n_j^{-1} \sum_{i=1}^{n_j} Y_{ji}, \qquad \bar{f}_{n,j} = \frac{1}{n_j} \sum_{i=1}^{n_j} f(W_{ji})
\]
and 
\[
\widehat{\var}_j(f(W)) = \frac{1}{n_j+N_j-1} \sum_{i=1}^{n_j+N_j}(f(W_{ji})-\bar{f}_j)^2, \qquad \bar{f}_j = \frac{1}{n_j+N_j} \sum_{i=1}^{n_j+N_j} f(W_{ji}).
\]
Defining
\begin{align*}
\widehat{\cov}_j(Y,f^*(W))  = \frac{1}{n_j-1} \sum_{i=1}^{n_j} (Y_{ji}-\bar{Y}_{j\cdot})(f^*(W_{ji})-\bar{f}^*_{n,j})
\end{align*}
and
\begin{align*}
\widehat{\var}_j(f^*(W))  = \frac{1}{n_j+N_j-1} \sum_{i=1}^{n_j+N_j}\left(f^*(W_{ji})-\bar{f}^*_j\right)^2
\end{align*}
for 
\[
\bar{f}^*_{n,j} = \frac{1}{n_j} \sum_{i=1}^{n_j} f^*(W_{ji}), \qquad \bar{f}^*_j = \frac{1}{n_j+N_j} f^*(W_{ji})
\]
as the usual sample covariances and variances,
since sample covariances and variances of i.i.d. samples are consistent for their population counterparts
we know that
\[
\lambda_j^* = \frac{N_j}{n_j+N_j}\frac{\widehat{\cov}_j(Y,f^*(W))}{\widehat{\var}_j(f^*(W))} \stackrel{p}{\rightarrow} \lambda_j
\]
in view of the fact that
\[
\frac{N_j}{n_j+N_j} = \frac{N_j/n_j}{1+N_j/n_j} \rightarrow \frac{1/\rho}{1+1/\rho} = \frac{1}{1+\rho}
\]
as $N_j \rightarrow \infty$ since $n_j/N_j \rightarrow \rho$.
So it remains to show that $\hat{\lambda}_j-\lambda_j^* = o_p(1)$,
or equivalently that 
\[
\widehat{\cov}_j(Y,f(W)) - \widehat{\cov}_j(Y, f^*(W)) = o_p(1) \qquad \text{ and } \widehat{\var}_j(f(W)) - \widehat{\var}_j(f^*(W)) = o_p(1).
\]

For the covariance we apply the Cauchy-Schwarz inequality to bound
\begin{align*}
\frac{n_j-1}{n_j}  \left|\widehat{\cov}_j(Y,f(W)) - \widehat{\cov}_j(Y,f^*(W))\right| & = \frac{1}{n_j} \Bigg| \sum_{i=1}^{n_j} (Y_{ji} - \bar{Y}_{j\cdot})(f^*(W_{ji})-\bar{f}^*_{n,j}-f(W_{ji})+\bar{f}_{n,j}) \Bigg|  \\
& \leq \sqrt{\frac{1}{n_j} \sum_{i=1}^{n_j}(Y_{ji}-\bar{Y}_{ji})^2} \times \sqrt{\frac{1}{n_j} \sum_{i=1}^{n_j}[(f^*(W_{ji})-f(W_{ji}))-(\bar{f}_{n,j}^*-\bar{f}_{n,j})]^2}.
\end{align*}
The first term 
\[
\sqrt{\frac{1}{n_j} \sum_{i=1}^{n_j}(Y_{ji}-\bar{Y}_{ji})^2}
\]
is essentially the sample standard deviation of the $Y_{ji}$ in the labeled dataset and hence of order $O_p(1)$ since $Y_{ji}$ is assumed to have finite variance.
We write the squared second term as
\begin{equation}
\label{eq:cov_expansion}
\frac{1}{n_j} \sum_{i=1}^{n_j}[(f^*(W_{ji})-f(W_{ji}))-(\bar{f}_{n,j}^*-\bar{f}_{n,j})]^2 = \frac{1}{n_j} \sum_{i=1}^{n_j} (f^*(W_{ji})-f(W_{ji}))^2 - (\bar{f}_{n,j}^*-\bar{f}_{n,j})^2.
\end{equation}
Using~\eqref{eq:f_diff_lbl} we see that 
\begin{align}
\frac{1}{n_j} \sum_{i=1}^{n_j} (f^*(W_{ji})-f(W_{ji}))^2 & = \hat{\beta}_1^2 \frac{1}{n_j} \sum_{i=1}^{n_j} \left(g(V_{ji})-\hat{g}^{(-k(j,i))}(W_{ji})\right)^2 \nonumber \\
+ &(\hat{\beta}-\beta)^{\top} \frac{1}{n_j}  \sum_{i=1}^{n_j} \psi^*(W_{ji})\psi^*(W_{ji})^{\top} (\hat{\beta}-\beta) \nonumber \\
 - &2\hat{\beta}_1 (\hat{\beta}-\beta)^{\top}\frac{1}{n_j} \sum_{i=1}^{n_j} \left(g(V_{ji})-\hat{g}^{(-k(j,i))}(V_{ji})\right)\psi^*(W_{ji}). \label{eq:lbl_squared_expansion}
\end{align}
But
\begin{align*}
\frac{1}{n_j} \sum_{i=1}^{n_j} \left(g(V_{ji})-\hat{g}^{(-k(j,i))}(W_{ji})\right)^2  = \sum_{k=1}^K \frac{n_j^{(k)}}{n_j} \frac{1}{n_j^{(k)}} \sum_{i:(j,i) \in \ci_k} \left(g(V_{ji})-\hat{g}^{(-k)}(W_{ji})\right)^2 = o_p(1)
\end{align*}
by~\eqref{eq:g_hat_vs_g},
and by Cauchy-Schwarz this implies
\begin{align*}
\left\| \frac{1}{n_j} \sum_{i=1}^{n_j} \left(g(V_{ji})-\hat{g}^{(-k(j,i))}(V_{ji})\right)\psi^*(W_{ji})\right\|  \leq \sqrt{\frac{1}{n_j} \sum_{i=1}^{n_j} \left(g(V_{ji})-\hat{g}^{(-k(j,i))}(V_{ji})\right)^2} \sqrt{\frac{1}{n_j} \sum_{i=1}^{n_j} \|\psi^*(W_{ji})\|_2^2}  = o_p(1)
\end{align*}
since the square integrability of $g(W_{ji})$ and $X_{ji}$ implies $\e[\|\psi^*(W)\|_2^2] < \infty$ and so in particular
\[
\frac{1}{n_j} \sum_{i=1}^{n_j} \|\psi^*(W_{ji})\|_2^2 = O_p(1).
\]
This fact along with the assumption that $\hat{\beta}-\beta = o_p(1)$ enable us to conclude that
\begin{equation}
\label{eq:f_star_vs_f_lbl}
\frac{1}{n_j} \sum_{i=1}^{n_j} (f^*(W_{ji})-f(W_{ji}))^2 = o_p(1)
\end{equation}
in view of~\eqref{eq:lbl_squared_expansion}.
Furthermore
\[
|\bar{f}_{n,j}^* - \bar{f}_{n,j}| = \left| \frac{1}{n_j} \sum_{i=1}^{n_j} f^*(W_{ji})-f(W_{ji})\right| \leq \sqrt{\frac{1}{n_j} \sum_{i=1}^{n_j} \left(f^*(W_{ji})-f(W_{ji})\right)^2} = o_p(1) 
\]
which by~\eqref{eq:cov_expansion} completes the argument that
\[
\widehat{\cov}_j(Y,f(W)) - \widehat{\cov}_j(Y, f^*(W)) = o_p(1).
\]

For the variance we use Cauchy-Schwarz again to write
\begin{align}
\frac{n_j+N_j-1}{n_j+N_j} \left|\widehat{\var}_j(f(W)) - \widehat{\var}_j(f^*(W)) \right| & \leq \frac{1}{n_j+N_j} \sum_{i=1}^{n_j+N_j} \left|(f(W_{ji})-\bar{f}_j)^2 - (f^*(W_{ji})-\bar{f}^*_j)^2\right| \nonumber \\
& \leq \sqrt{\frac{1}{n_j+N_j} \sum_{i=1}^{n_j+N_j} (f(W_{ji})-\bar{f}_j+f^*(W_{ji})-\bar{f}^*_j)^2} \nonumber \\
& \times \sqrt{\frac{1}{n_j+N_j} \sum_{i=1}^{n_j+N_j} \left(f(W_{ji})-f^*(W_{ji})-(\bar{f}_j-\bar{f}_j^*)\right)^2}. \label{eq:var_cauchy_schwarz}
\end{align}
First we show that
\begin{equation}
\label{eq:var_second_term}
\frac{1}{n_j+N_j} \sum_{i=1}^{n_j+N_j} \left(f(W_{ji})-f^*(W_{ji})-(\bar{f}_j-\bar{f}_j^*)\right)^2 = o_p(1).
\end{equation}
Similar to~\eqref{eq:cov_expansion} we can expand the left-hand side of~\eqref{eq:var_second_term} as
\begin{equation}
\label{eq:var_expansion}
\frac{1}{n_j+N_j} \sum_{i=1}^{n_j+N_j} \left(f(W_{ji})-f^*(W_{ji})-(\bar{f}_j-\bar{f}_j^*)\right)^2 = \frac{1}{n_j+N_j} \sum_{i=1}^{n_j+N_j} \left(f^*(W_{ji})-f(W_{ji})\right)^2  - (\bar{f}^*_j-\bar{f}_j)^2.
\end{equation}
We have
\begin{align}
\frac{1}{n_j+N_j}  \sum_{i=1}^{n_j+N_j} \left(f^*(W_{ji})-f(W_{ji})\right)^2 & = \frac{n_j}{n_j+N_j} \frac{1}{n_j} \sum_{i=1}^{n_j}\left(f^*(W_{ji})-f(W_{ji})\right)^2  + \frac{N_j}{n_j+N_j} \frac{1}{N_j} \sum_{i=n_j+1}^{n_j+N_j} \left(f^*(W_{ji})-f(W_{ji})\right)^2 \label{eq:lbl_unlbl_weighted_avg}
\end{align}
and by~\eqref{eq:f_diff_unlbl} we know that
\begin{align}
\frac{1}{N_j} \sum_{i=n_j+1}^{n_j+N_j} \left(f^*(W_{ji})-f(W_{ji})\right)^2 & = \hat{\beta}_1^2 \frac{1}{N_j} \sum_{i=n_j+1}^{n_j+N_j} \left(g(V_{ji})-\frac{1}{K} \sum_{k=1}^K \hat{g}^{(-k)}(V_{ji})\right)^2 \nonumber + \\&(\hat{\beta}-\beta)^{\top} \frac{1}{N_j}\sum_{i=n_j+1}^{n_j+N_j} \psi^*(W_{ji})\psi^*(W_{ji})^{\top} (\hat{\beta}-\beta) \nonumber - \\&2\hat{\beta}_1(\hat{\beta}-\beta)^{\top} \frac{1}{N_j}\sum_{i=n_j+1}^{n_j+N_j} \left(g(V_{ji})-\frac{1}{K}\sum_{k=1}^K \hat{g}^{(-k)}(V_{ji})\right)\psi^*(W_{ji}). \label{eq:unlabeled_squared_expansion}
\end{align}
We have
\begin{align*}
\frac{1}{N_j} \sum_{i=n_j+1}^{n_j+N_j} \left(g(V_{ji})-\frac{1}{K} \sum_{k=1}^K \hat{g}^{(-k)}(V_{ji})\right)^2  & = \frac{1}{K^2N_j} \sum_{i=n_j+1}^{n_j+N_j} \left(\sum_{k=1}^K g(V_{ji})-\hat{g}^{(-k)}(V_{ji})\right)^2 \\
& = \frac{1}{K^2N_j} \sum_{i=n_j+1}^{n_j+N_j}\sum_{k=1}^K\sum_{\ell=1}^K \left(g(V_{ji})-\hat{g}^{(-k)}(V_{ji})\right)\left(g(V_{ji})-\hat{g}^{(-\ell)}(V_{ji})\right).
\end{align*}
Since
\[
\frac{1}{N_j} \sum_{i=n_j+1}^{n_j+N_j} \left(g(V_{ji})-\hat{g}^{(-k)}(V_{ji})\right)^2 = o_p(1), \quad k=1,\ldots,K
\]
by an identical argument to the proof of~\eqref{eq:g_hat_vs_g},
using Cauchy-Schwarz we conclude
\begin{align*}
\bigg| \frac{1}{N_j} \sum_{i=n_j+1}^{n_j+N_j}\left(g(V_{ji})-\hat{g}^{(-k)}(V_{ji})\right)\left(g(V_{ji})-\hat{g}^{(-\ell)}(V_{ji})\right)\bigg| 
&  \leq \sqrt{\frac{1}{N_j} \sum_{i=n_j+1}^{n_j+N_j} \left(g(V_{ji})-\hat{g}^{(-k)}(V_{ji})\right)^2 } \\
&  \quad+ \sqrt{\frac{1}{N_j} \sum_{i=n_j+1}^{n_j+N_j} \left(g(V_{ji})-\hat{g}^{(-\ell)}(V_{ji})\right)^2 } \\
&  = o_p(1)
\end{align*}
for each $k,\ell \in \{1,\ldots,K\}$ and so
\[
\frac{1}{N_j} \sum_{i=n_j+1}^{n_j+N_j} \left(g(V_{ji})-\frac{1}{K} \sum_{k=1}^K \hat{g}^{(-k)}(V_{ji})\right)^2 = o_p(1).
\]
One more application of Cauchy-Schwarz then gives
\begin{align*}
\left\| \frac{1}{N_j} \sum_{i=n_j+1}^{n_j+N_j} \left(g(V_{ji})-\frac{1}{K}\sum_{k=1}^K \hat{g}^{(-k)}(V_{ji})\right)\psi^*(W_{ji})\right\| & \leq \sqrt{\frac{1}{N_j} \sum_{i=n_j+1}^{n_j+N_j} \left(g(V_{ji})-\frac{1}{K}\sum_{k=1}^K \hat{g}^{(-k)}(V_{ji})\right)^2} \\
&\quad \times \sqrt{\frac{1}{N_j} \sum_{i=n_j+1}^{n_j+N_j} \|\psi^*(W_{ji})\|_2^2} \\
& = o_p(1)
\end{align*}
Reminding ourselves again that $\hat{\beta}-\beta = o_p(1)$ by assumption enables us to conclude
\[
\frac{1}{N_j} \sum_{i=n_j+1}^{n_j+N_j} \left(f^*(W_{ji})-f(W_{ji})\right)^2 = o_p(1)
\]
by~\eqref{eq:unlabeled_squared_expansion} and then that
\[
\frac{1}{n_j+N_j} \sum_{i=1}^{n_j+N_j} \left(f^*(W_{ji})-f(W_{ji})\right)^2 = o_p(1)
\]
by~\eqref{eq:f_star_vs_f_lbl} and~\eqref{eq:lbl_unlbl_weighted_avg}.
This immediately implies
\[
|\bar{f}_j^*-\bar{f}_j| \leq \left|\frac{1}{n_j+N_j} \sum_{i=1}^{n_j+N_j} f^*(W_{ji})-f(W_{ji})\right| \leq \sqrt{\frac{1}{n_j+N_j} \sum_{i=1}^{n_j+N_j}\left(f^*(W_{ji})-f(W_{ji})\right)^2} = o_p(1)
\]
and so~\eqref{eq:var_second_term} holds, as desired.
Finally
\begin{align*}
\sqrt{\frac{1}{n_j+N_j} \sum_{i=1}^{n_j+N_j} (f(W_{ji})-\bar{f}_j+f^*(W_{ji})-\bar{f}^*_j)^2} &  \leq \sqrt{\frac{1}{n_j+N_j} \sum_{i=1}^{n_j+N_j} \left(f(W_{ji})+f^*(W_{ji})\right)^2}  \\
&  \leq \sqrt{\frac{1}{n_j+N_j} \sum_{i=1}^{n_j+N_j} (f(W_{ji}))^2}  + \sqrt{\frac{1}{n_j+N_j} \sum_{i=1}^{n_j+N_j} (f^*(W_{ji}))^2} \\&
\quad\quad\text{ by Minkowski's inequality} \\
&  \leq 2\sqrt{\frac{1}{n_j+N_j} \sum_{i=1}^{n_j+N_j} (f(W_{ji}))^2} \\&\quad+ \sqrt{\frac{1}{n_j+N_j} \sum_{i=1}^{n_j+N_j} (f^*(W_{ji})-f(W_{ji}))^2} \\&
\quad\quad\text{ by Minkowski's inequality again} \\
&  = O_p(1) \text{ by square integrability of $f^*(W_{ji})$}
\end{align*}
which completes the proof by~\eqref{eq:var_cauchy_schwarz}.
\end{proof}

\end{document}